\documentclass[%
preprint,
 amsmath,amssymb,
 aps,
]{revtex4-2}

\usepackage{graphicx}
\usepackage{dcolumn}
\usepackage{bm}

\usepackage[dvipsnames]{xcolor}
\usepackage{todonotes}

\begin{document}

\preprint{APS/123-QED}



\title{Driven transitions between megastable quantized orbits}

\author{Álvaro G. López$^1$}\email{alvaro.lopez@urjc.es}
\author{Rahil N. Valani$^2$}%
 
\affiliation{$^1$Nonlinear Dynamics, Chaos and Complex Systems Group.\\Departamento de F\'isica, Universidad Rey Juan Carlos, Tulip\'an s/n, M\'ostoles,28933, Madrid,Spain}

\affiliation{$^2$Rudolf Peierls Centre for Theoretical Physics, Parks Road,
University of Oxford, OX1 3PU, United Kingdom}

\date{\today}

\begin{abstract}
We consider a nonlinear oscillator with state-dependent time-delay that displays a countably infinite number of nested limit cycle attractors, \emph{i.e.} megastability. In the low-memory regime, the equation reduces to a self-excited nonlinear oscillator and we use averaging methods to analytically show quasilinear increasing amplitude of the megastable spectrum of quantized quasicircular orbits. We further assign a mechanical energy to each orbit using the Lyapunov energy function and obtain a quadratically increasing energy spectrum and (almost) constant frequency spectrum. We demonstrate transitions between different quantized orbits, i.e. different energy levels, by subjecting the system to an external finite-time harmonic driving. For large driving amplitude with frequency close to the limit cycle frequency, resonance drives transitions to higher energy levels. Alternatively, for large driving amplitude with frequency slightly detuned from limit-cycle frequency, beating effects can lead to transitions to lower energy levels. Such driven transitions between quantized orbits form a classical analog of quantum jumps.
For excitations to higher energy levels, we show amplitude locking where nearby values of driving amplitudes result in the same response amplitude, i.e. the same final higher energy level. We rationalize this effect based on the basins of different limit cycles in phase space. From a practical viewpoint, our work might find applications in physical and engineering system where controlled transitions between several limit cycles of a multistable dynamical system is desired.
\end{abstract}

\maketitle

\section{Introduction}

Energy quantization is largely considered an exclusive feature of microscopic quantum mechanical systems. Classical Hamiltonian systems that conserve energy possess a continuous spectrum of energy values. However, there are examples of classical dynamical systems that display many quantized orbits as limit cycles, and hence a discrete energy spectrum, when oscillatory non-conservative forces that result in dissipation and self-excitation are present~\cite{lopez2024mega,Kahn2014-na,ZARMI201721}. Real-life systems, such as mechanical and electrodynamic systems, always dissipate energy due to friction forces, which ultimately relies on some electrodynamic radiative process. 
A potential source of self-excitation in dynamical systems is time-delay feedback. Delays frequently arise as a consequence of the finite speed of information propagation. They are pertinent when a medium or a field produces the forces between two interacting bodies~\cite{raju2004electrodynamic} and when long causal chains in a connected network are reduced in a model. Time-delayed feedback mechanisms have been demonstrated in many disciplines of science, ranging from complex chemical reactions \cite{schell1986effects}, to mechanical physical systems \cite{airy1830certain}, or complex biological systems \cite{mackey1977oscillation,hansen2022effect,ferrell2011modeling}. They are also important in climate phenomena \cite{boutle2007nino}, epidemiology, and population dynamics \cite{salpeter1998mathematical}.

Inspired from electrodynamics of moving bodies, a state-dependent delay system with a retarded harmonic potential has been shown to exhibit a dynamical analog of orbit quantization, with a finite set of quantized orbits as limit cycles~\cite{LOPEZ2023113412}. Moreover, numerical simulations as well as experiments with silicone oil walking droplets demonstrate a hydrodynamic quantum analog of quantized orbits in a harmonic potential, due to the effects of time-delay-induced self-forces~\citep{Couder2005,Perrard2014a,Bush2020review}. In the literature on dynamical systems, the existence of a countably infinite structure of stable periodic orbits is known as \emph{megastability}. It was first numerically reported for periodically driven nonlinear oscillators~\cite{Sprott2017}. Inspired by this, in a recent work, we found that a dynamical analog of a countably infinite number of quantized orbits via megastability can be shown in minimal models of walking droplets and electrodynamically-inspired delay systems~\cite{lopez2024mega,Lopez2024selfexcited}. In this spirit of exploring dynamical analogs of quantum pheonomena using minimal dynamical systems, in the present work we explore transitions between megastable quantized orbits as a dynamical analog of quantum jumps. To the best of authors' knowledge, there have been limited studies of transitions between several stable orbits~\cite{grines2018describing,4252155}, and none from the point of view of a classical analog of quantum jumps. Moreover, the results presented in this manuscript might also find practical use in engineering dynamical systems with multiple limit cycles where controlled transitions between them are desired.


The paper is organized as follows. In Sec.~\ref{Sec: DS} we introduce the state-dependent delay system describing a retarded harmonic oscillator. By considering the low-memory regime of the system, in Sec.~\ref{Sec: OQ} we use averaging methods to obtain analytical approximations of the megastable spectrum of limit cycle orbits. We further characterize the energy spectrum of orbits in Sec.~\ref{Sec: ES}. In Sec.~\ref{sec: DT}, we present our main results of transitions between megastable orbits via a finite-time harmonic pulse and characterize their nature in detail. Section~\ref{sec: RC} is devoted to study the resonance curves. Lastly, we discuss and conclude in Sec.~\ref{Sec: Discussion and Conclusion}.

\section{Retarded harmonic oscillator model}\label{Sec: DS}

Consider an inertial particle of mass $m$, immersed in an external harmonic potential $V(x)= k x^2/2$ and experiencing a linear drag force with damping coefficient $\mu$. The particle is also subjected to a self-force, which is represented via a retarded harmonic potential~\cite{lien98,onanelec} $Q(x_{\tau})=\alpha x_{\tau}^2/2$ having state-dependent time-delay, where $x_{\tau}=x(t-\tau(\dot{x}))$. According to Newton's second law, the dynamics of this system is governed by the equation of motion
\begin{equation}
    m\ddot{x}+\zeta \dot{x} + \dfrac{dV}{dx}+\dfrac{dQ}{dx_{\tau}} = 0. 
    \label{eq:1}
\end{equation}
This state-dependent dynamical system 
has previously been shown to exhibit several coexisting attractors (both limit cycles and chaotic attractors) for a certain choice of $\tau({x})$~\cite{lopval24}. The time-delayed force is non-conservative and it makes the system self-excited i.e. the system can gain and lose energy and exhibit limit-cycle oscillations~\cite{jenkins}. 
The amplitude $\alpha$ controls the strength of the delay force on the particle. We choose the delay function to take the form $\tau(\dot{x})=\tau_0 \cos^{2}(\lambda \dot{x})$. The dependence of time-delay on particle velocity is inspired by the Liénard-Wiechert potentials \cite{onanelec}, while the oscillatory nature of delay is inspired from models of wave-driven walking droplets in the low-memory regime~\citep{lopez2024mega}. Moreover, this functional dependence of time-delay on velocity ensures limited memory and reflection spatial symmetry (\emph{i.e.} Eq.~\eqref{eq:1} is invariant under the transformation $x \rightarrow -x$). The parameter $\tau_0$ represents the maximum value of the memory, which is attained when the particle velocity is zero, at the turning points. Dividing by $m$ and rescaling $x\rightarrow x/\lambda$, ${\zeta}\rightarrow \zeta/m$, $k\rightarrow k/m$ and ${\alpha}\rightarrow\alpha/m$ results in the following equation
\begin{align}
\ddot{x}  + {\zeta} \dot{x} + {k} x + {\alpha} x_{\tau}=0, 
\label{eq:2}
\end{align}
with $x_\tau=x(t-\tau(\dot{x}))$ and $\tau(\dot{x})=\tau_0 \cos^2\dot{x}$. This rescaling is not intended to modify the units of the variables, but only its values. The dynamical system shown in Eq.~\eqref{eq:2} can be numerically solved using a residual control integrator~\cite{shampine05}. For simulations presented in this paper, we consider constant initial history functions $x(t)=x_0$ for $t<0$ where $x_0$ is a constant. This choice of history function comprise a subset of initial condition in the true infinite-dimensional phase space of the delay system.

\section{Quantization of orbits as megastability}\label{Sec: OQ}
We analyze the system in Eq.~\eqref{eq:2} in the low-memory regime and show that the equation reduces to a nonlinear self-sustained oscillator. 
In the low-memory regime corresponding to small $\tau_0$, a Taylor series expansion of the time-delay term to first order~\cite{Erneu23} transforms Eq.~\eqref{eq:2} to the following
\begin{equation}
     \ddot{x}+ (\zeta-\alpha \tau_0 \cos^2\dot{x})\dot{x} + (k+\alpha) x = 0.    
    \label{eq:3}
\end{equation}
Equation~\eqref{eq:3} describes a nonlinear self-excited oscillator with an oscillatory nonlinear drag term and a rescaled harmonic potential. Introducing $\omega^2=k+\alpha$, $\mu=\zeta-\alpha \tau_0/2$ and $\epsilon=\alpha \tau_0/2$ results in the following nonlinear oscillator
\begin{equation}
    \ddot{x} + (\mu - \epsilon \cos\dot{x})\dot{x} + x= 0, 
    \label{eq:4}
\end{equation}
where we have rescaled $(x,t,\epsilon,\mu) \rightarrow (x/2\omega, t/\omega, \epsilon \omega, \mu \omega)$ to further simplify the equation. In the low-memory limit and the weak dissipation regime ($\zeta \approx \alpha \tau_0/2$), we can assume that $\mu$ and $\epsilon$ are small and apply the Krylov-Bogoliubov averaging method~\cite{krylov1950} to obtain two coupled ordinary differential equations governing the dynamics of the amplitude and the phase of the limit cycles in the phase space of the system. In the phase space, we rewrite our dynamical system as
\begin{align}\label{eq:5}
\dot{x} & = y \\ \nonumber
\dot{y} & = - (\mu-\epsilon \cos y) y - x.
\end{align}
 \begin{figure}
\centering
\includegraphics[width=0.7\columnwidth]{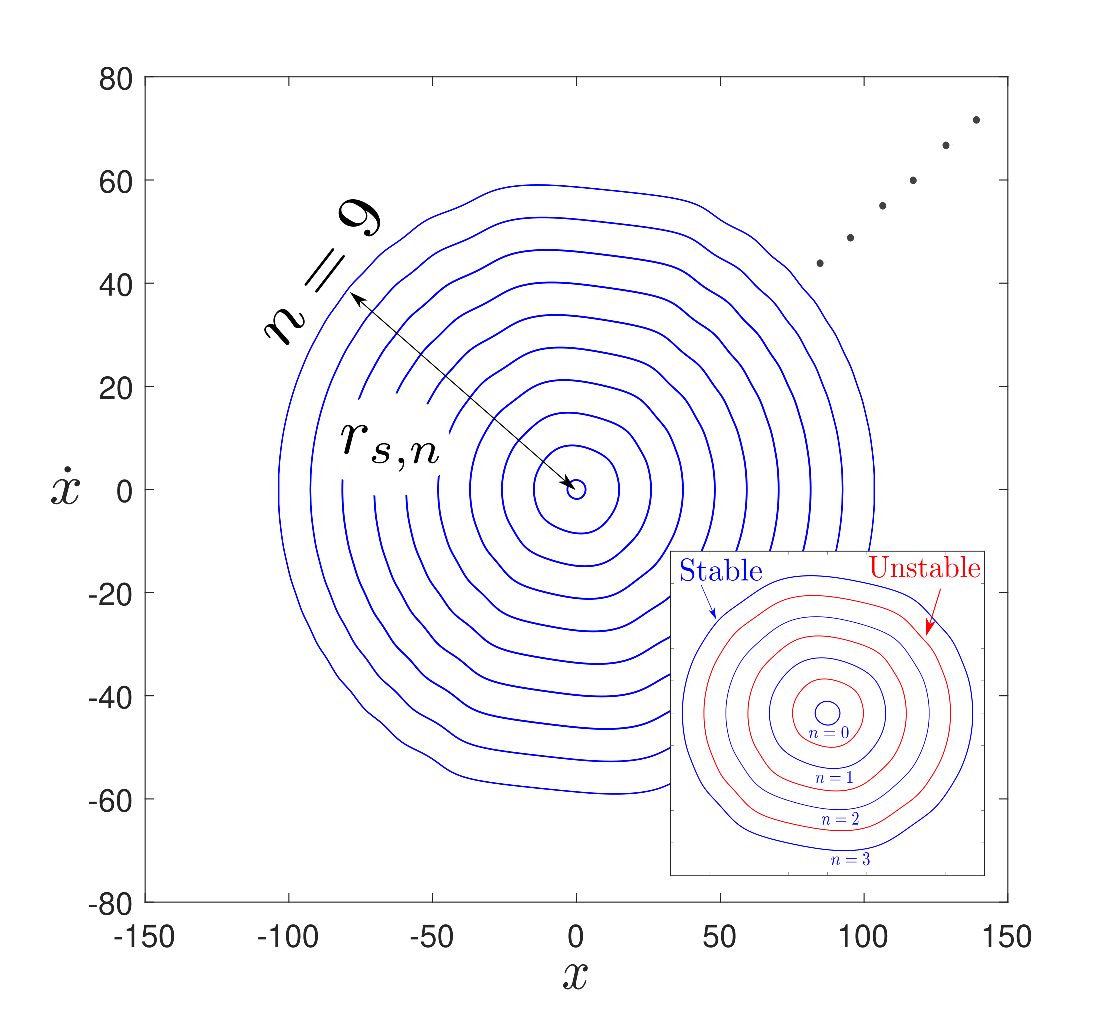}
\caption{A countably infinite megastable set of quantized orbits, i.e. stable limit cycles, obtained by numerically solving Eq.~\eqref{eq:2} (dots indicate that the sequence continues). In the inset we show the first four stable limit cycles (blue), separated by their unstable counterparts (red), which correspond to the basin boundaries. The parameter values are fixed to $k=1/10$, $\alpha=1/4$, $\zeta=1/10$, $\lambda=1/2$, $\tau_0=4/5$ giving $\epsilon=1/10$ and $\mu=0$.}
\label{Fig:1}
\end{figure}

The Krylov-Bogoliubov averaging method proposes the phase space ansatz $(x(t),y(t))=(r(t) \sin(t+\varphi(t)),r(t) \cos(t+\varphi(t)))$, and averages over the phase variable $\theta(t) = t+\varphi(t)$, invoking the fact that for small $\mu$ and $\epsilon$, a time-scale separation exists between the variations of the phase $\theta(t)$ and the amplitude $r(t)$ of the oscillation. The error of this approximation is bounded and can be made arbitrarily small by decreasing the size of $\max(\varepsilon,\mu)$ \cite{krylov1950}. Substituting the ansatz in Eq.~\eqref{eq:5}, we obtain the following system of differential equations for the amplitude and the phase
\begin{align}
\dot{r} & = - r (\mu-\epsilon \cos (r \cos \theta)) \cos^2 \theta. \label{eq:61} \\
\dot{\varphi} & = (\mu-\epsilon \cos (r \cos \theta)) \cos \theta \sin \theta. 
\label{eq:62}
\end{align}
Averaging Eqs.~\eqref{eq:61} and \eqref{eq:62} over the period $\theta \in [0,2\pi]$ of the unperturbed ($\mu=\epsilon=0$) harmonic oscillator yields $\dot{\varphi}=0$ and
\begin{equation}
\dot{r}  = -  \mu r+\epsilon (J_{1} (r)- r J_{2}(r)),
\label{eq:7}
\end{equation}
where $J_{n}(r)$ is the $n$th order Bessel function of the first kind. In the particular case that $\mu=0$, the asymptotic expansion of the Bessel function $J_{n}(r) \approx \sqrt{\pi/2r} \cos(r- n \pi/2-\pi/4)$ allows to approximately find the fixed points of Eq.~\eqref{eq:7} as the roots of the equation $\tan(r-3\pi/4)=1/r$. Since the hyperbola has a horizontal asymptote at zero and the tangent function is strictly periodic, we have the existence of a countably infinite number of equilibria, with alternating stability. The series of zeros converge to the zeros of the tangent function, thus obeying the asymptotic relation $r_n \approx \pi(3/4 + n)$, where $r_n$ denotes the $n$th root of Eq.~\eqref{eq:7}. Thus, we obtain an infinite nested structure of limit cycles with alternating stability that gives rise to megastability for the nonlinear oscillator.
 \begin{figure}
\centering
\includegraphics[width=0.5\columnwidth]{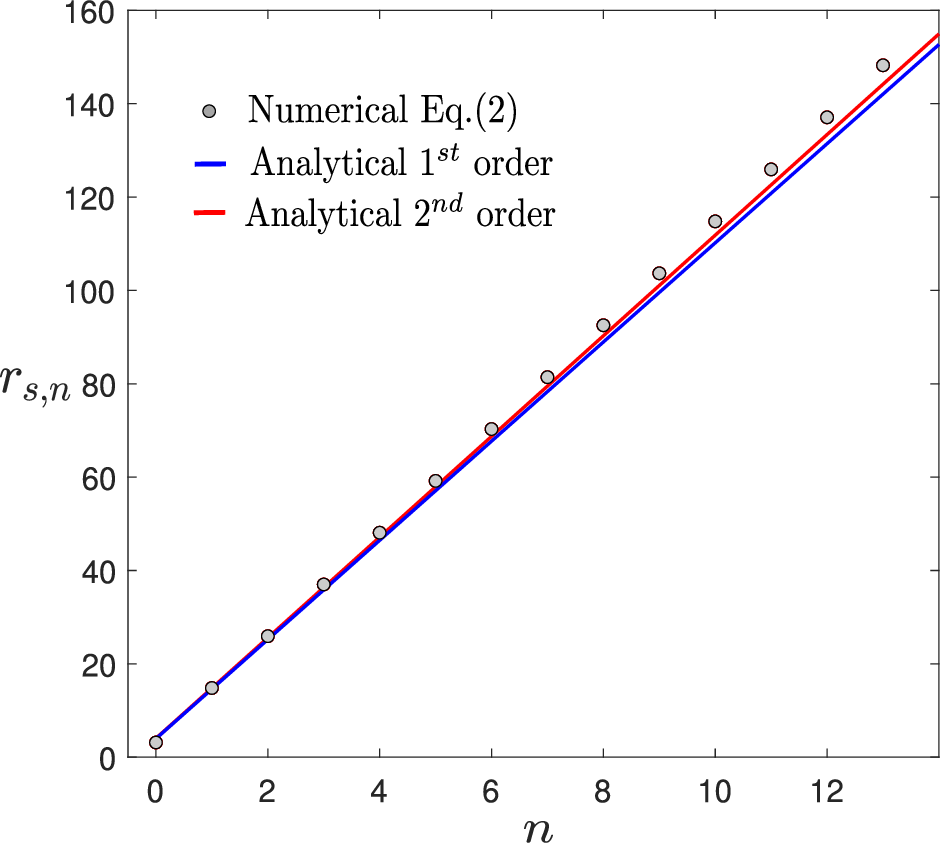}
\caption{Dependence of the radius $r_{s,n}$ of the stable limit cycle orbits on the orbit number $n$. Grey dots represent the numerical results obtained by numerically integrating Eq.~\eqref{eq:2}, with the same parameter values as in Fig.~\ref{Fig:1}. The blue line represents the value predicted by the analytical Krylov-Bogoliubov approximation, considering a first order in the Taylor series expansion of the state-dependent time-delay. The the red line results from considering the same approximation, but to second order in the delay, which includes averaging $\tau^2(\dot{x})$ along the periodic orbits. This results in a renormalized value of the mass $m+3\alpha \tau_0^2/16$.}
\label{Fig:1R}
\end{figure}
 
To compare our analytical approximations in the low-memory regime with the state-dependent delay system in Eq.~\eqref{eq:1}, we perform numerical simulations. In Fig.~\ref{Fig:1} we have computed the limit cycles of Eq.~\eqref{eq:1} beyond the first ten, using constant history functions $x(t)=x_0$ for $t<0$, with increasing values of $x_0$. The parameter values $k=1/10$, $\alpha=1/4$, $\zeta=1/10$, $\lambda=1/2$ and $\tau_0=4/5$ are fixed hereafter. These parameters give $\epsilon=1/10$ and $\mu=0$. Numerical simulations agree well with the analytical predictions of the location of the orbits and they also confirm that the size of limit cycles within the megastable structure continues to increase without limit. As can be seen in Fig.~\ref{Fig:1R}, the limit cycle size increases following an asymptotically linear trend with $n$, although the analytical estimation given by the Krylov-Bogoliubov method underestimates the limit cycle size. 

In the more general case $\mu>0$, there is only a finite number $N_c$ of limit cycles. Solving for $\dot{r}=0$, and using again the asymptotic expansion of the Bessel functions, we obtain the transcendental equation $r=\cos(r-3 \pi/4)/(\sqrt{2 r/\pi}\mu/\varepsilon+\sin(r-3 \pi/4))$. For $r \gg 1$, the maximum number of limit cycles is restricted to $N_c=\lfloor \delta/2 \pi-3/8\rfloor$, where $\delta=(\pi/2)^{1/3}(\epsilon/\mu)^{2/3}$. This gives us a scaling relation for the number of stable limit cycles in the form $N_c(\epsilon,\mu) \propto (\epsilon/\mu)^{2/3}$. As the dissipation coefficient $\mu$ increases in the interval $\left[0, \infty \right)$, an infinite sequence of subcritical saddle-node bifurcations, in decreasing order of orbit size, annihilates the megastable structure of periodic orbits.
 \begin{figure}
\centering
\includegraphics[width=0.7\columnwidth]{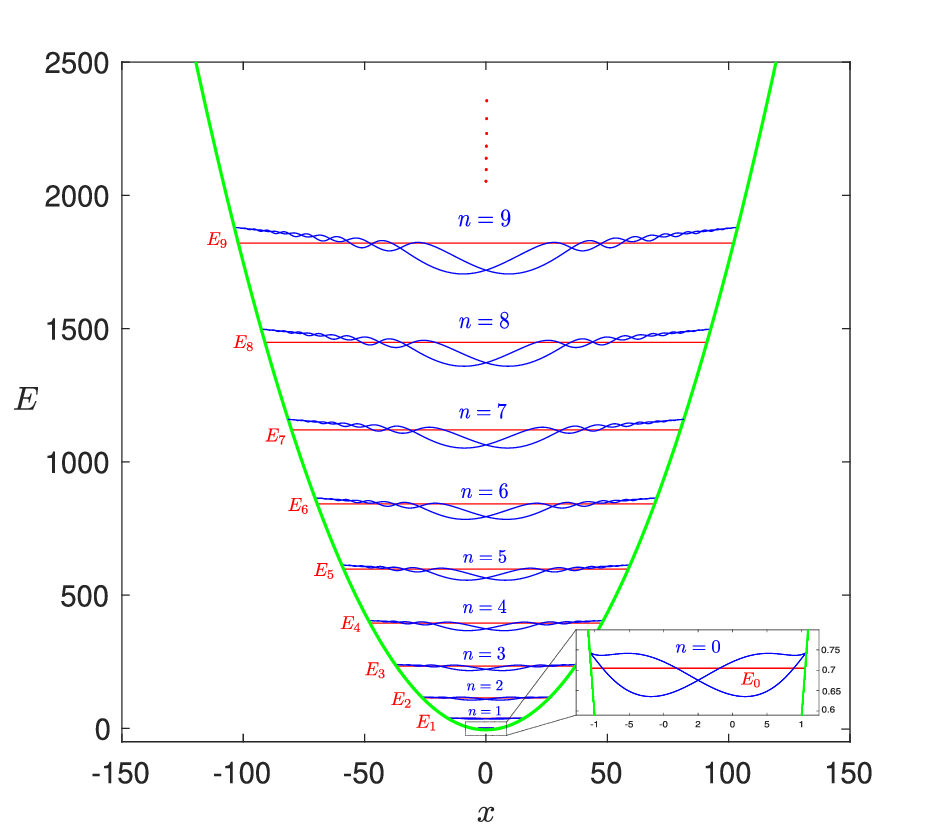}
\caption{The first ten mean Lyapunov energy levels $E_n$ (red) in the external harmonic well potential (green) for the megastable orbits in Fig.~\ref{Fig:1}, together with the time evolution of the Lyapunov energy function $E(x(t),y(t))$ (blue) for orbits with different levels $n$.}
\label{Fig:2}
\end{figure}

\section{Energy spectrum of quantized orbits}\label{Sec: ES}

To characterize the nature of the energy spectrum, we compute both the energy and frequency spectra along quantized orbits. Following recent studies~\cite{onanelec} connecting self-excited oscillators and the quantum potential~\cite{bohm52}, we define the energy of our oscillator using the Lyapunov function associated with Eq.~\eqref{eq:1}, when $\tau_0=0$ and also $\mu=0$. This function describes the mechanical energy content of the oscillator, which is conserved when self-excitation and dissipation are neglected. It corresponds to the first perturbative contribution to the energy of the oscillator \cite{Erneu23}, and can be written as \footnote{Note, we could have used $k$ instead of $k+\alpha$, regarding to the original Eq.~\eqref{eq:1}, as well. However, we have considered Eq.~\eqref{eq:3}, instead. A careful consideration hints at the use of the latter, since the delay term in Eq.~\eqref{eq:1} includes a conservative component when this term is expanded in a Taylor series, and it relates to the zeroth-order contribution (see Ref.~\cite{Davidow2017}). Thus, we use $k+\alpha$, which gives more accurate fitting results.} 
\begin{equation}
E(x,y) = \frac{1}{2} m y^2 + \frac{1}{2} (k+\alpha) x^2.
\label{eq:8_1}
\end{equation}

Hamiltonian dynamical systems always maintain a constant value of the mechanical energy $E(x,y)$ and require fine-tuned parameters for dissipation and self-excitation, $\mu=0$ and $\epsilon=0$, respectively. In contrast, self-sustained oscillators gain and lose energy during different parts of the oscillation, forming a thermodynamic engine~\cite{lopezte}. The non-conservative terms in Eq.~\eqref{eq:2} implies that mechanical energy is gained by the oscillator during some part of the limit cycle, while mechanical energy is lost during the complementary part. This fact is related to the time-reversal asymmetry of Eqs.~\eqref{eq:2} and~\eqref{eq:4}, and, more generally, of self-excited systems \cite{jenkins,mackey2011}. The resulting dissipative structure~\cite{Pri78} constitutes an open dynamical system, which conserves its energy \emph{on average}. Certainly, as explained in previous works~\cite{lopez2024mega,Davidow2017}, the computation of the Melnikov function for Eq.~\eqref{eq:2} or Eq.~\eqref{eq:3} is equivalent to averaging the equations for the radial component of the limit cycle, which shows that the Melnikov function is zero along the megastable quantized orbits. Thus, the Lyapunov energy function appearing in Eq.~\eqref{eq:8_1} is conserved on average along the limit cycles corresponding to Eq.~\eqref{eq:2}. 

\begin{figure}
\centering
\includegraphics[width=0.6\columnwidth]{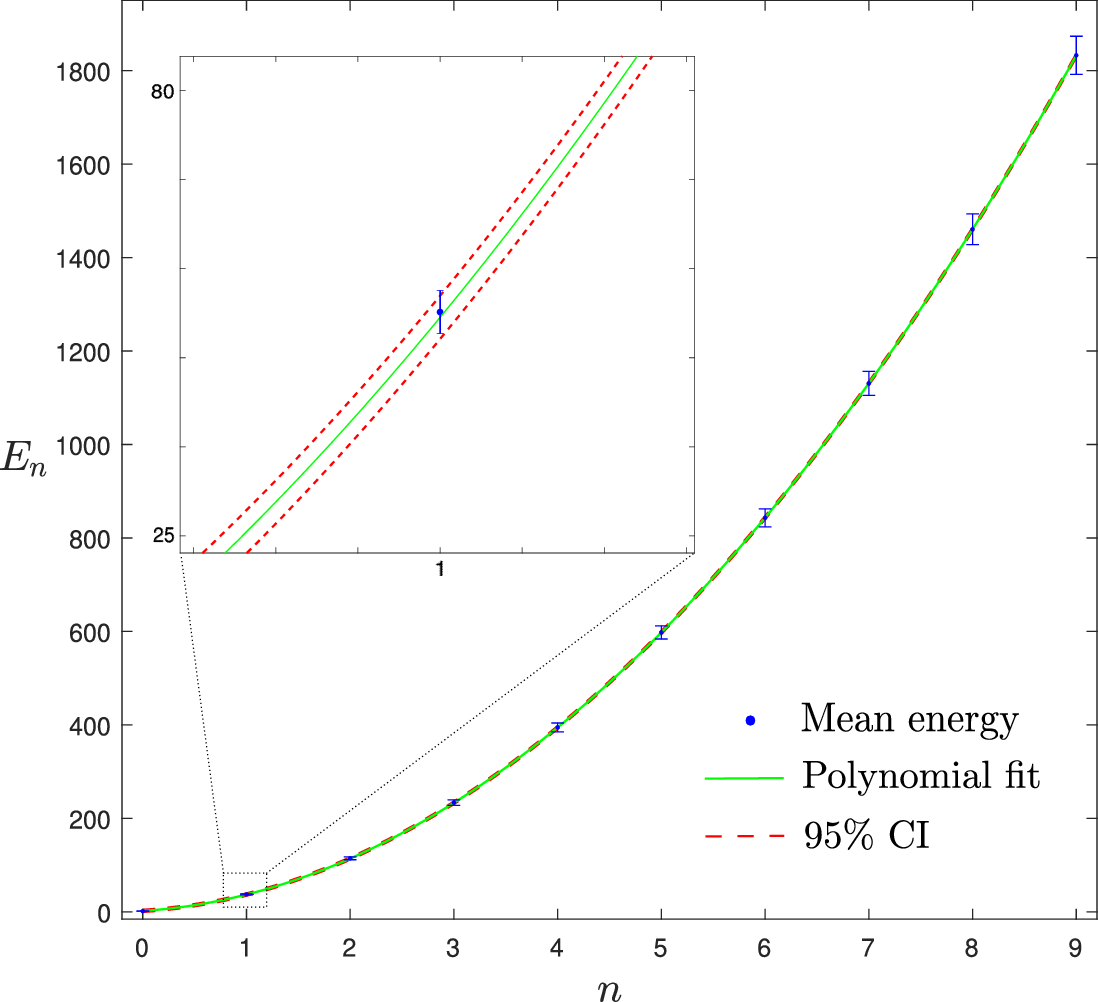}
\caption{The discrete spectrum of energy levels in Fig.~\ref{Fig:2} follow a quadratic trend. The least-squares fitting is very accurate, with tight 95 \% confidence intervals (CI). The error bars represent the energy uncertainty along the quantized orbit, computed as the quadratic deviation of the Lyapunov energy function from its average value.}
\label{Fig:3}
\end{figure}

Consequently, we assign to each stable limit cycle in Fig.~\ref{Fig:1} a mean energy defined as the time average along the limit cycle of the Lyapunov energy function $E_n=\int^T_0 E(x_n(t),y_n(t))d t/T$ where $(x_n(t),y_n(t))$ represents the trajectory of the $n$th limit cycle. These energy levels are defined in the external harmonic potential with $k+\alpha$, and are shown in Fig.~\ref{Fig:2}. Since the stable limit cycles appearing in Eq.~\eqref{eq:1} are approximately circular, we approximate their size as $r_{s,n} \approx \pi (3/4 + 2 n)/2\lambda$ according to Eq.~\eqref{eq:2}.  This equation justifies our previous choice of $\lambda=1/2$, for simplicity. Thus we can estimate the spectrum of Eq.~\eqref{eq:5} as $E_n \approx r^2_{s,n}/2 = 2 \pi^2 (3 /8 + n)^2 $, obtaining a parabolic trend for the mean  energy as a function of the orbit number $n$. This trend is confirmed in Fig.~\ref{Fig:3}, where a least-squares fitting to the quantization rule $E_n=a n^2+b n+c$ gives the parameter values $a=21.04$, $b=13.95$ and $c=2.27$. The parameter $a$ correlates with the analytical result $2 \pi^2$, while the two remaining parameters $b=3 \pi^2/2$ and $c=9 \pi^2/32$ are also close. The differences are attributable to the distortion introduced by higher-order terms in our delayed system (see Ref.~\cite{lopez2024mega} for comparison), distorting the first quantized energy levels and making the limit cycle not strictly quasiharmonic (see again Fig.~\ref{Fig:1}).

From our analysis, the period of the limit cycles remains almost constant for the entire megastable set. Thus, the frequency spectrum should not vary with the orbit number, and this is confirmed by numerical simulations. The Fourier analysis of the periodicity of the spectrum of the limit cycle set is quasimonochromatic with an average asymptotic value of $\omega_n=0.59$, nearly coinciding with the predicted value of $\omega_n=\sqrt{k+\alpha}$ from our low-memory equation. This constitutes a fundamental difference with respect to quantum mechanical systems, for which both the energy and the frequency obey a linear relation with quantum number $n$. Nevertheless, we suggest the possibility that different energy-frequency quantum-like functional relations may arise in our system of other nonlinear confining potentials $V(x)$ and time-delay functions $\tau(\dot{x})$. The averaging Krylov-Bogoliubov method that we have used for the harmonic potential can be easily extended to any one-dimensional integrable Hamiltonian system, allowing us to engineer to some extent the energy-frequency spectrum of these megastable structures. 

\section{Driven transitions between quantized orbits}\label{sec: DT}

We now study transitions between quantized orbits by driving our oscillator with a periodic pulse having a finite duration of time $\delta t$. Several types of nonlinear resonances can occur in our system, depending on the frequency detuning between the natural frequency of our self-excited oscillator and the frequency of the driving. More specifically, we show that both excitations from lower energy levels to high energy levels and jumps from high energy levels to low energy levels are possible.
\begin{figure}
\centering
\includegraphics[width=0.7\columnwidth]{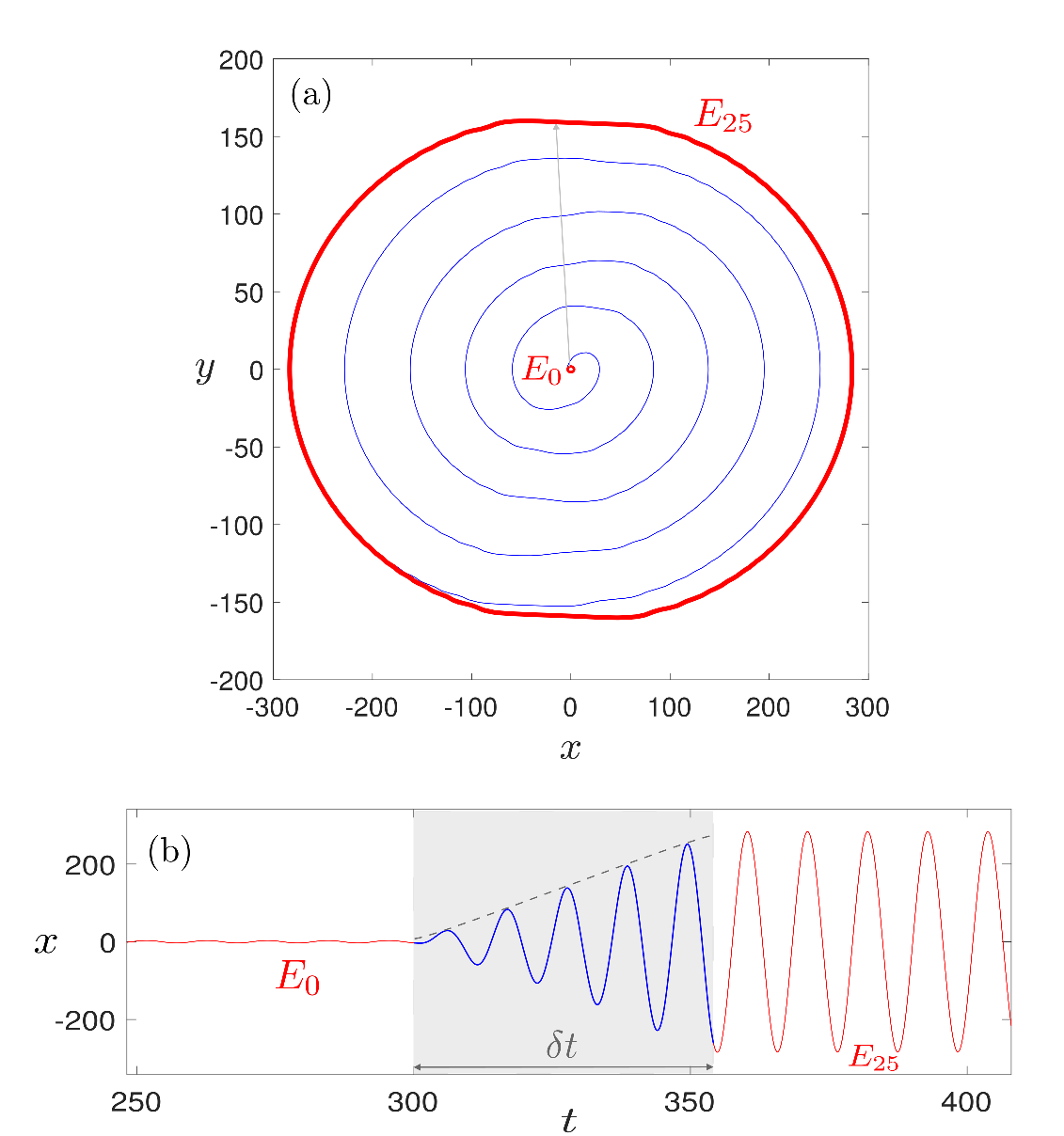}
\caption{An excitation of the self-oscillator from the zero-point energy level $E_0$ to a very high energy level $E_{25}$ (red curves), under the application of a finite-duration harmonic pulse (blue curve and gray shaded region) from $t_0=300$ with parameters $F_0=6.0$, $\Omega=0.59$ and $\delta t=2 \pi N/\Omega$, with $N=5$. Other parameter values are same as Fig.~\ref{Fig:1}, except for $\tau_0=0.82$, which has been slightly increased. (a) An Archimedean spiral (blue) in phase space representing the transition between two limit cycles (red). (b) Time series of the transition, where the window (gray) represents the duration of the pulse $\delta t$, that produces a resonant response.}
\label{fig:4}
\end{figure}

We choose our driving function $F(t)$ as follows
\[ F(t)=\begin{cases} 
      F_0 \cos(\Omega t + \phi), & \text{if}\:\:t_0 \, \leq t \leq t_0+\delta t, \\
      0, & \text{otherwise}.  
   \end{cases}.
\]
This function allows us to simulate the effect of a periodic pulse of varying amplitude, frequency, initial phase and time window on our self-oscillating system appearing in Eq.~\eqref{eq:1}. Unless otherwise stated, we consider the time of initiation of the harmonic pulse to be $t_0=300$, which is large enough for transients to decay and the system to settle on a particular quantized orbit. We also fix $\phi=0$, since variations in the initial phase of the driving force have a relatively insignificant effect on the transitions between orbits compared to the effects of varying $F_0$, $\Omega$ and $\delta t$. Finally, we slightly increase the delay feedback from $\tau_0=4/5$ to $\tau_0=0.82$, to ensure that our previous results are robust against small perturbations in $\tau_0$, without any other significant consequences. 

In the limit of small amplitudes $F_0$ and frequencies close to the natural frequency of the limit cycles, the phenomenon of synchronization takes place~\citep{Pikovsky_Rosenblum_Kurths_2001}. During synchronization, the phase of the self-excited oscillator along the limit cycle orbit is locked to the phase of the driving pulse, whereas the amplitude of the limit cycle hardly changes. Thus, no transitions take place between limit cycle orbits. However, as the driving amplitude $F_0$ is increased, deviation from the original limit cycle takes place and the concept of {resonance} is more appropriate to describe the dynamics of our driven self-excited oscillator. This eventually leads to the system transitioning to a new limit cycle after the driving pulse ceases. 

In Fig.~\ref{fig:4} we show the effect of a long-lived pulse whose frequency is finely tuned to the natural frequency of the limit cycles $\omega_n\approx 0.59$. 
As we can see in Fig.~\ref{fig:4}(b), in this case the amplitude of the oscillator grows linearly in time, reminiscent of a conventional linear resonance phenomenon. In the phase space picture shown in Fig.~\ref{fig:4}(a), this translates into an approximately Archimedean spiral starting from the fundamental energy level $E_0$ and leading to the excited orbit $E_{25}$. By increasing the magnitude of the forcing $F_0$, such a high orbit number may also be reached with a pulse of shorter duration (see Sec.~\ref{sec: RC}). Thus, in this classical resonance regime, when the driving frequency is close to that of the natural frequency of the oscillator, only transitions from lower to higher megastable orbits can be realized.

\begin{figure}
\centering
\includegraphics[width=0.7\columnwidth]{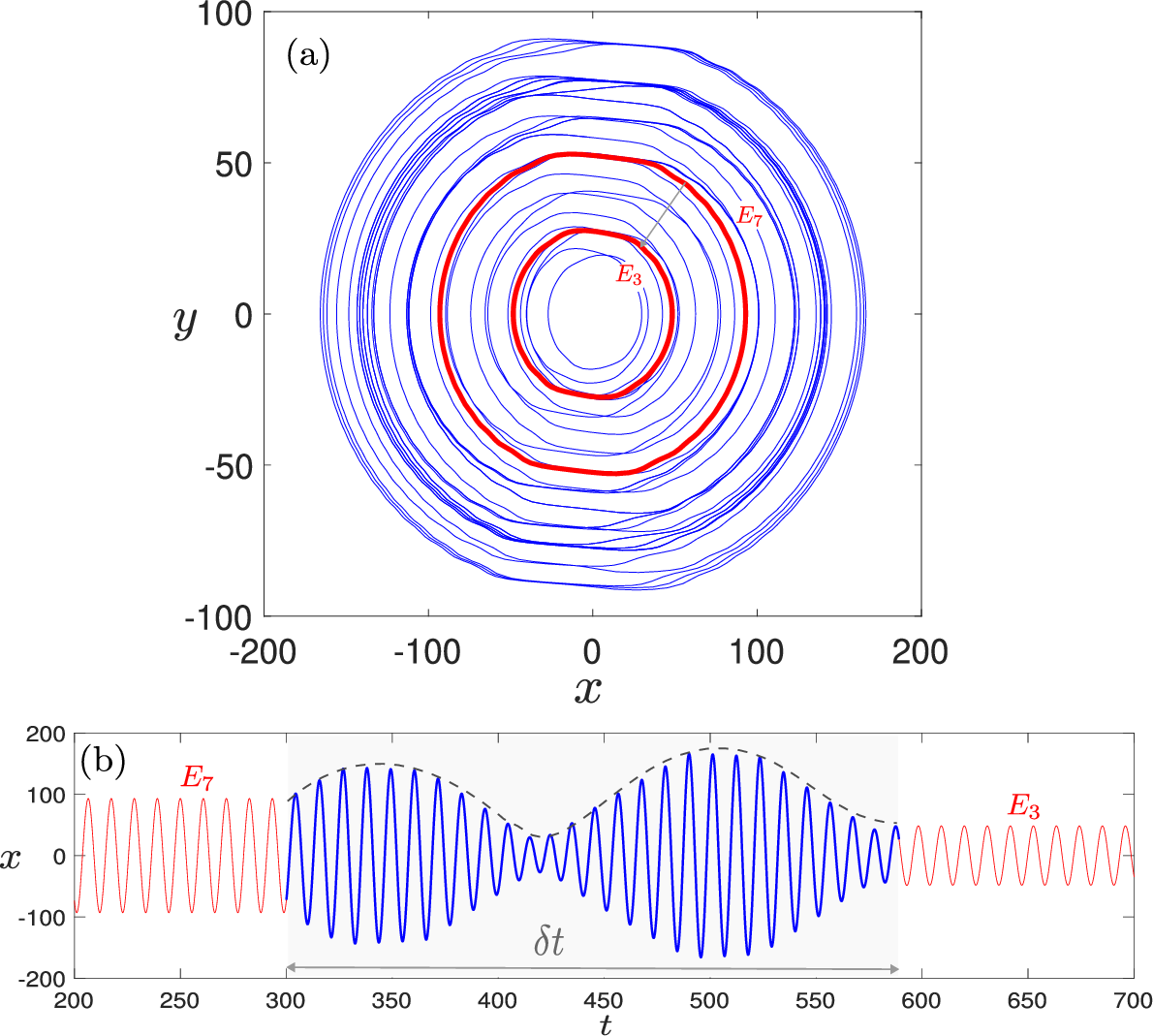}
\caption{A transition of the self-excited oscillator from a higher energy level $E_7$ to a lower energy level $E_3$ (red curves) taking place by forcing via a finite-duration harmonic pulse (blue curve and gray shaded region) initiated at $t_0=300$ with parameters $F_0=3.0$, $\Omega=0.54$ and $\delta t=2 \pi N/\Omega$, with $N=25$. Other parameter values are same as Fig.~\ref{Fig:1} except for $\tau_0=0.82$. (a) A quasiperiodic spiral (blue) in phase space sweeping a wide range of limit cycles. (b) Time series of the transition, where the window (gray) represents the duration of the pulse $\delta t$, producing a beating response.}
\label{Fig:5}
\end{figure}

By slightly detuning the driving frequency from the natural frequency of the self-oscillator, for example, by setting $\Omega=0.54$, the oscillator can enter a beating mode, and we obtain more complex oscillations with a periodically modulated amplitude as the number of driving periods $N$ becomes sufficiently large. This amplitude modulation allows the self-oscillator to cross over a wider range of megastable orbits both above and below the starting energy level. 
Thus, with detuned frequencies, we can not only transition to higher energy levels, but also it is possible to make transitions to low energy levels. The specific energy level obtained at the end of the driving pulse will depend on the basin of attraction in which the oscillator finds itself at the end of the pulse. An example of such a transition from a higher to a lower energy level ($E_7$ to $E_3$) is shown in Fig.~\ref{Fig:5}. 

Hence, we see that by appropriately engineering a harmonic pulse of finite duration, we can controllably transition between different quantized orbits of our megastable spectrum. We expect these results to hold for other self-excited megastable systems, as long as the basins of attraction of the nested limit cycles are smoothly separated. However, we also note that as the detuning is increased even further, transient quasiperiodic motion and even transient chaotic dynamics (\emph{e.g.} for $\Omega=1.0$) are observed for the duration of the pulse. The existence of transient chaos gives rise to unpredictability in the final orbit that the system will settle into after the end of the pulse. 

\section{Resonance curves} \label{sec: RC}

To gain further insight into the dependence of transitions on the driving pulse parameters, we have characterized the effects of variations in the parameters $\Omega$, $F_0$ and $\delta t$ on the response amplitude $Q$
\cite{gandhimathi2006vibrational}. 
This variable $Q$ allows us to capture the asymptotic response of the system to an external forcing, depending on its frequency, amplitude, and duration. The response of a system to an external perturbation can be decomposed as a set of frequency components
\begin{equation}
x(t)=\sum_{i=0}^\infty Q(\omega_i)\cos(\omega_i t+\phi_i),
\label{eq:9_1}
\end{equation}
where $\omega_i$ are the frequencies and $\phi_i$ the phases contained in the response, while $Q(\omega_i)$ is the amplitude response of each frequency component. Decomposing the series into its even and odd components, we can define the amplitudes
\begin{equation}
Q_c(\omega) = \frac{2}{n T}\int_{t_a}^{t_a+n T} x(t) \cos(\omega t) d t, ~~~ Q_s(\omega) =  \frac{2}{n T} \int_{t_a}^{t_a+ n T}x(t) \sin(\omega t) d t,
\label{eq:8}
\end{equation}
and the response amplitude $Q_t(\omega)=\sqrt{Q^2_c(\omega)+Q^2_s(\omega)}$, which can be regarded as the Fourier transform of a particular trajectory $x(t)$ at a specified frequency $\omega$ in the window $n T$, where here $n$ is a fixed number of cycles and $T$ is some fixed characteristic periodicity of the system, which in our system corresponds to the period of the limit cycles $2\pi/\omega_n$. When a driving force is applied steadily, $t_a$ represents the time at which the driving force begins to act. However, here the time $t_a$ represents an instant for which we can consider that the trajectory has reached the final megastable orbit, and thus the transition has been completed (thus $t_a>t_0+\delta t$). Unless otherwise stated, we fix $t_a=500$. Finally, we consider the maximum value $Q=\text{max}\{Q_t(\omega)\}$ over the entire spectrum of frequencies $\omega$, which is located close to the spatial amplitude of the limit cycles, given their quasiharmonicity.

\begin{figure}
\centering
\includegraphics[width=0.6\columnwidth]{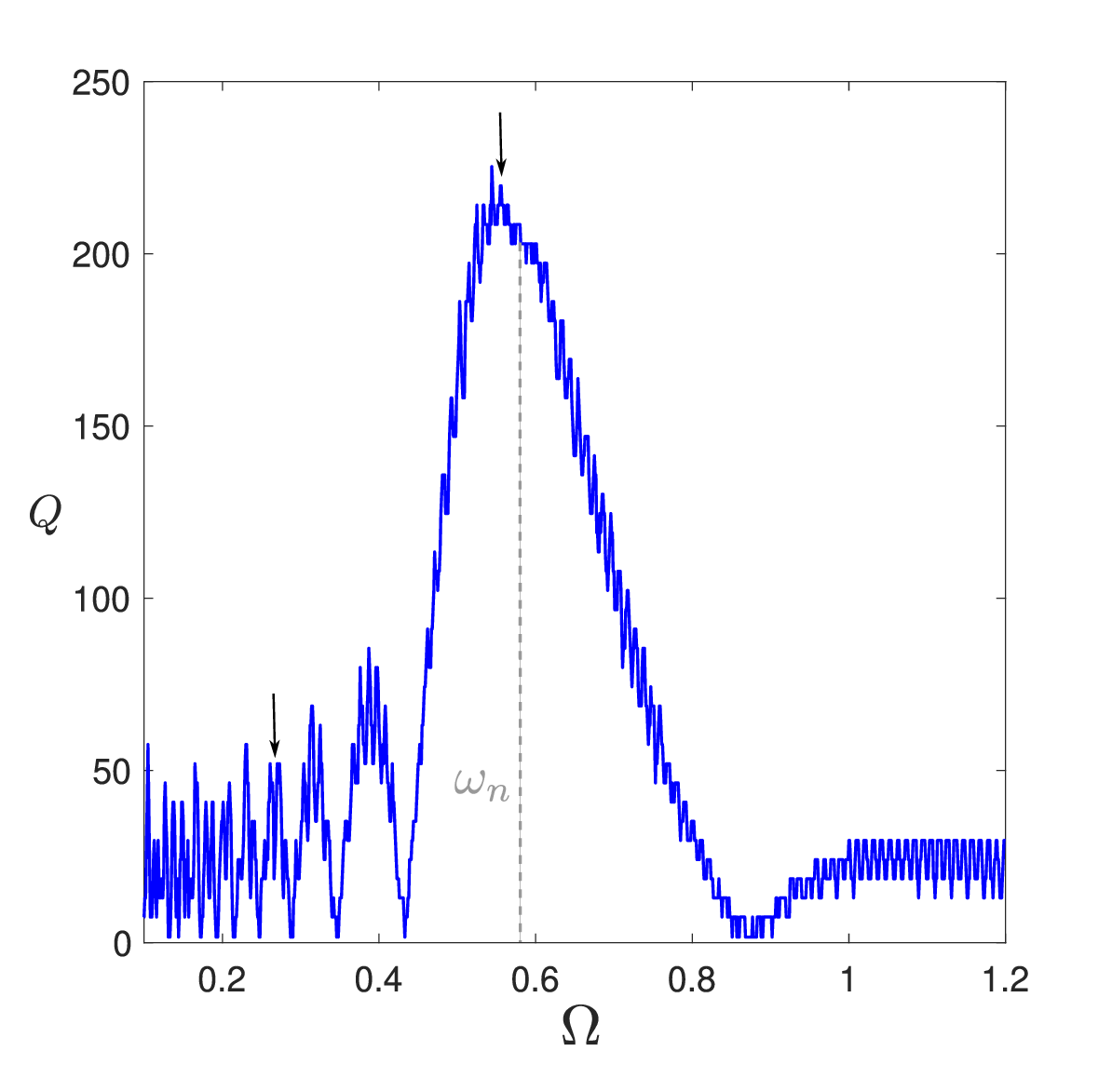}
\caption{Resonance curve representing the amplitude response of the system $Q$ as a function of the driving frequency, for $N=3$ and $F_0=15$. We see a peak at the resonant frequency, very close to the frequency of the megastable orbits $\omega_n$. Secondary peaks can be identified at lower frequencies, among which we find the first subharmonic $\omega_n/2$ (black arrows). Other parameter values are same as Fig.~\ref{Fig:1} except for $\tau_0=0.82$.}
\label{fig:56}
\end{figure}

It is well known that an underdamped forced harmonic oscillator under the influence of an (endless) harmonic periodic perturbation shows a peak in the response amplitude very close to the natural frequency. The amplitude of oscillation at the resonant frequency is directly proportional to the driving amplitude. Here, since our oscillator is nonlinear and we are applying finite-time perturbations, we encounter a dynamically richer scenario. As shown in Fig.~\ref{fig:56}, when the response amplitude $Q$ is represented against the frequency of the driving $\Omega$, for the system starting at the fundamental zero-point energy level $E_0$, a pronounced peak is observed near the frequency of the limit cycles, which can be approximated by $\omega_n=0.59$. However, unlike linear systems, we also observe secondary peaks at lower frequencies, whose frequencies cannot be identified as particular rational fractions of the limit cycle frequencies, except for the subharmonic $\omega_n/2$.
Moreover, it is interesting to note that this nonlinear resonance curve is not smooth but consists of discrete irregular jumps, due to the megastable sets being at a finite distance apart and also because we are using perturbations lasting a particular time window $\delta t$. 

\begin{figure}
\centering
\includegraphics[width=0.6\columnwidth]{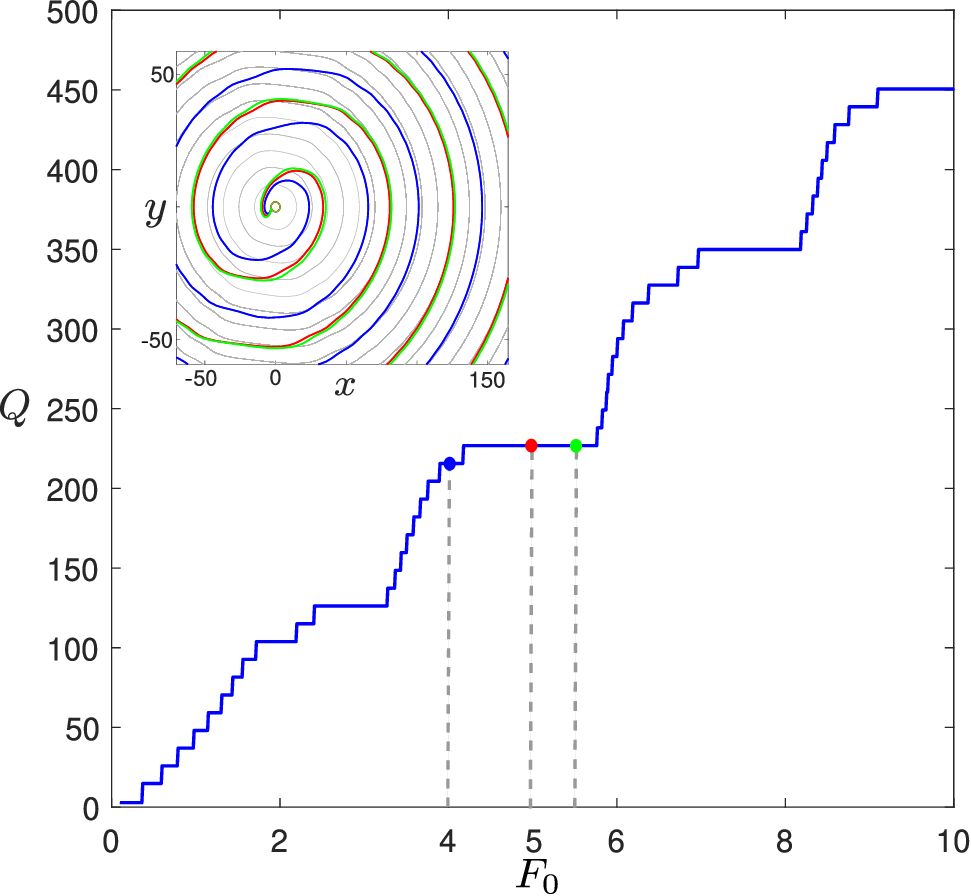}
\caption{Response curve showing the amplitude $Q$ as a function of the perturbation amplitude $F_0$. The response amplitude increases with the driving amplitude on average. However, a sequence of amplitude plateaus are detected, that arise due to attraction of trajectories by the megastable set in the phase space. The inset shows three particular trajectories (blue, red and green) corresponding to three different $F_0$ values (see colored dots). The parameter $N=5$ and $\Omega=0.58$. Other parameter values are same as Fig.~\ref{Fig:1} expect for $\tau_0=0.82$.}
\label{fig:6}
\end{figure}

We computed the system response $Q$ as a function of the driving amplitude $F_0$ for a fixed frequency near the resonant frequency, again the system starting at the fundamental zero point energy level $E_0$, with parameter values $N=5$ and at the resonant peak $\Omega=0.58$. The resulting resonance curve is quite different from the conventional linear dependence between the response amplitude and the amplitude of the driving force. Despite the fact that on average we see an increase in response amplitude with increasing driving amplitude, as shown in Fig.~\ref{fig:6}, we also observe a sequence of plateaus showing amplitude locking and rendering a very complicated dependence between both variables. These complicated nonlinear functional relation can be rationalized in terms of the underlying megastable structure in the phase space. 

As can be seen in the inset of Fig.~\ref{fig:6}, during the initial driving cycles, as trajectories with different $F_0$ spiral outwards and cross megastable orbits (especially the unstable ones, which lie in between the gray stable orbits depicted in the inset of Fig.~\ref{fig:6}), the trajectories can get pushed apart (blue and red) if they land on different sides of the unstable orbits  or the trajectories can get attracted (green and red) towards each other if they land on the same side of the unstable orbit i.e. in the same basin of attraction. Orbits that were initially pushed apart continue to follow different Archimedean spirals for the duration of the pulse and end up at different final energy levels. Conversely, orbits that were attracted initially end up following the same Archimedean spiral and end up at the same energy level at the end of the applied pulse. The latter yields plateaus in the response curve separated by regions of stepwise growth due to the former. Thus, it appears that the fate of the trajectories is decided by the megastable structure during the initial driving cycles. The fact that the megastable sets are regular is reflected in the regularity of plateaus. These effects are robust against changes in the phase $\phi$~\footnote{Variations in $\phi$ seems to introduce an uncertainty of one energy level during transitions i.e. the final orbit after harmonic pulse is above or below one level. In fact, the effect of modifying $t_0$ is very similar to changing the value of $\phi$, and both parameters have a weak effect on transitions between orbits.
}, and they persist as the duration of the pulse $\delta t$ decreases. Importantly, we highlight that this plateau sequence is enhanced when $\Omega \approx \omega_n$. This is expected since the nonlinear resonance that increases the orbital amplitude is most effective for driving near the natural frequency.


\begin{figure}
\centering
\includegraphics[width=0.7\columnwidth]{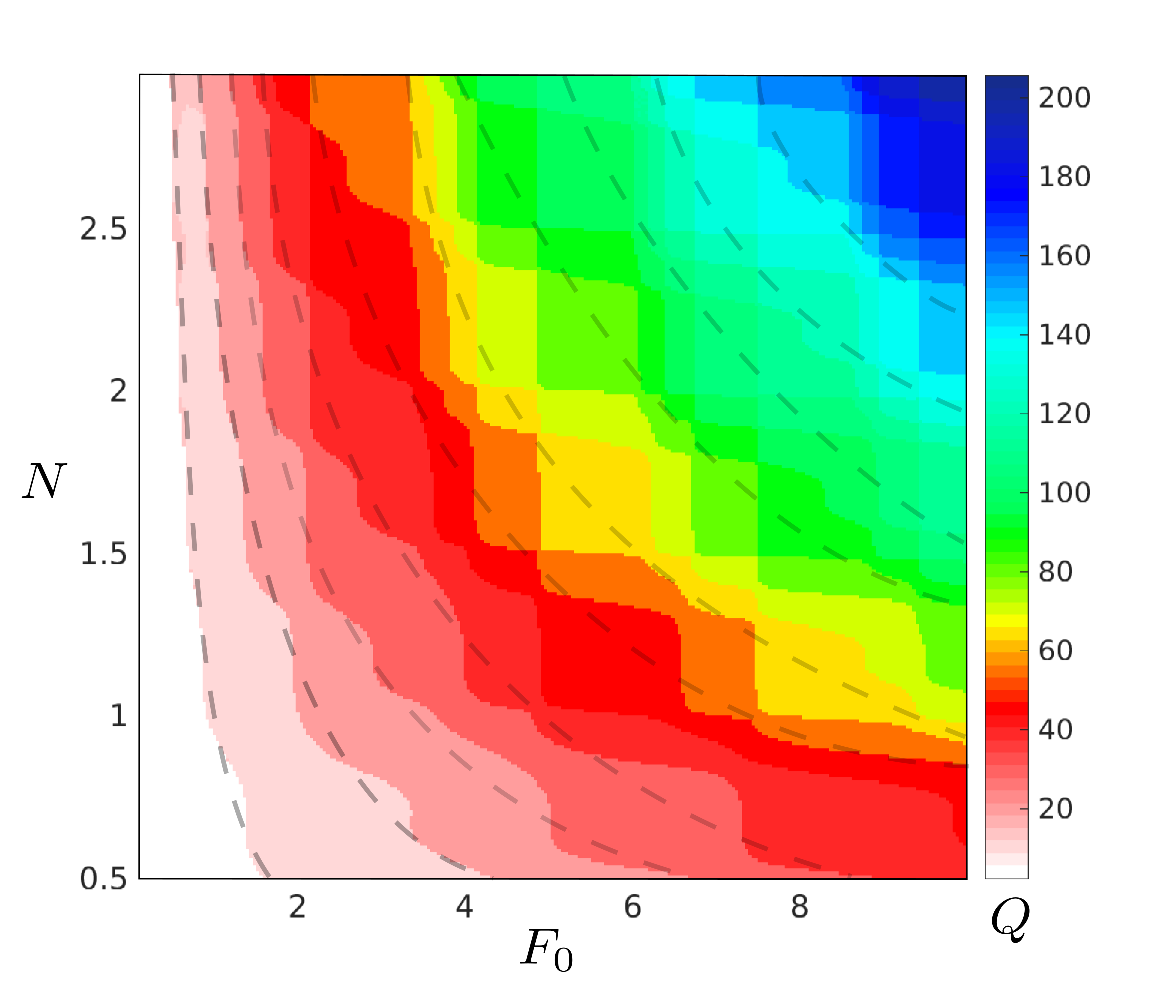}
\caption{Amplitude response $Q$ as a function of the driving amplitude $F_0$ and the number of driving cycles $N$. The driving frequency has been set to $\Omega=\omega_n$ and we recall that $\delta t=2 \pi N/\Omega$. To speed up simulations, we have reduced the time at which the forcing begins to $t_0=200$ and the arrival time to $t_a=400$. Other parameter values are same as Fig.~\ref{Fig:1} except for $\tau_0=0.82$. We can see that the duration of the pulses $N$ can be reduced as much as desired as long as the driving amplitude is correspondingly enhanced. The level curves mark the transitions between megastable orbits. The grey curves depict the approximate profile of level curves and are for guide only.}
\label{fig:7}
\end{figure}

Lastly, we have computed the amplitude response as a function of the driving amplitude $F_0$ and the number of cycles $N$ at the resonance frequency $\omega_0$, starting at the fundamental energy level. As previously stated, given some megastable orbit, one can reach any particular higher orbit by applying a pulse with certain amplitude over a sufficiently large time window. However, in some circumstances one might prefer to apply a very short but intense pulse to produce the same effect more quickly. As shown in the density plot appearing in Fig.~\ref{fig:7}, this is certainly possible. 
Regions with the same color correspond to values in the parameter space $(F_0,N)$ that reach the same megastable orbit. The level curves appearing in this plot separating these regions correspond to changes in the destination orbit. It is interesting to note that there is a general tendency of the level curves to decrease in $N$ as $F_0$ increases. If we consider moving in a non-inertial oscillating frame with the particle, it can be assumed that the driving force is approximately constant \emph{i.e.} $F_0$. Then, since the energy gained $\Delta E$ is proportional to impulse, which in turn is proportional to the duration of the pulse, it can be conjectured that the level curves of constant energy differences between orbits may follow a hyperbolic trend given by $\Delta E \sim F_0 N$. Nevertheless, from Fig.~\ref{fig:7} we can appreciate considerable undulations, and the landscape complicates noticeably as the driving amplitude and the number of driving cycles increase.

\section{Discussion and Conclusion}\label{Sec: Discussion and Conclusion}

Limit cycles abound in classical, nonlinear, self-excited dynamical systems, especially when time-delayed feedback is present, due to the ubiquity of the Hopf bifurcation \cite{jenkins, doi:10.1142/S0218127409025250}. Given the general assumptions of the present work, we expect that many dynamical systems of this kind to possess megastable structures, with corresponding infinite spectra of energy levels. 
In these megastable systems, orbit quantization implies a balance between energy gain (self-excitation) and energy loss (dissipation) from non-conservative forces during the orbit, so that the average Lyapunov energy function is conserved along their limit cycle trajectories. This picture, when naively compared with energy orbitals of atoms, contrasts with Bohr's belief that electrons do not emit electromagnetic radiation along quantized orbits \cite{bohr13}, and suggests an alternative dynamical viewpoint based on the balance between dissipation and self-excitations to understand the dynamics of the avoidance of atomic collapse~\cite{raju2004electrodynamic, Pen14}.

 
The stability of the megastable quantized states described in this paper can be characterized by the nature and size of their basins of attraction~\cite{nuss96}. The present work shows that the set of initial conditions leading to a certain limit cycle attractor is separated by unstable orbits, producing smooth basin boundaries (see Fig.~\ref{Fig:1}). Once the system has settled on a quantized orbit, an external impulse to the system can drive transitions between different limit cycles of this megastable structure, forming an analog to quantum jumps. We have demonstrated transitions both from low to high energy levels and vice versa,
in this countably infinite set of orbits. The two transitions involve different transient dynamics, with the former involving ordinary resonance, while the latter requires beating via slight detuning of the driving frequency from that of the limit-cycle orbits. Further, it is important to note that once the system settles to a particular attractor, any information about its initial condition is erased due to dissipation. Moreover, in systems where the limit sets are chaotic, fractal basins yield an asymptotically unpredictable system~\cite{LOPEZ2023113412}. These two considerations allows one to conceptualize a dynamical analog of collapse towards a quantum eigenstate after measurement, as a perturbed megastable dynamical system approaching a quantized attractor \cite{bohm52}.


Lastly, limit cycles are ubiquitous in many physical and engineering systems, such as electrical \cite{van1960theory}, mechanical \cite{DUNNMON20111182}, electromechanical \cite{PhysRevApplied.22.034072}, and optomechanical~\cite{xu2022limit}. Whenever coexisting nested limit cycles are present in such systems as a consequence of {self-excitation}, the formalism presented in this paper may be used to induce controlled transitions between limit cycles.
First, the phase of a limit cycle orbit can be synchronized (without switching to a different limit cycle) to the phase of an external driving force via small-amplitude forcing near the limit cycle frequency. Once the phase is controlled, the amplitude of the driving can be enhanced to take the system towards higher or lower limit cycles via resonances or beating effects, respectively, as described in this paper. This recipe might be useful in driving controlled transitions between different limit cycles in classically quantized multistable systems.  

\textit{Acknowledgments}. R.V. acknowledges the support of the Leverhulme Trust [Grant No. LIP-2020-014].

\bibliography{apssamp}

\begin{thebibliography}{43}%
\makeatletter
\providecommand \@ifxundefined [1]{%
 \@ifx{#1\undefined}
}%
\providecommand \@ifnum [1]{%
 \ifnum #1\expandafter \@firstoftwo
 \else \expandafter \@secondoftwo
 \fi
}%
\providecommand \@ifx [1]{%
 \ifx #1\expandafter \@firstoftwo
 \else \expandafter \@secondoftwo
 \fi
}%
\providecommand \natexlab [1]{#1}%
\providecommand \enquote  [1]{``#1''}%
\providecommand \bibnamefont  [1]{#1}%
\providecommand \bibfnamefont [1]{#1}%
\providecommand \citenamefont [1]{#1}%
\providecommand \href@noop [0]{\@secondoftwo}%
\providecommand \href [0]{\begingroup \@sanitize@url \@href}%
\providecommand \@href[1]{\@@startlink{#1}\@@href}%
\providecommand \@@href[1]{\endgroup#1\@@endlink}%
\providecommand \@sanitize@url [0]{\catcode `\\12\catcode `\$12\catcode `\&12\catcode `\#12\catcode `\^12\catcode `\_12\catcode `\%12\relax}%
\providecommand \@@startlink[1]{}%
\providecommand \@@endlink[0]{}%
\providecommand \url  [0]{\begingroup\@sanitize@url \@url }%
\providecommand \@url [1]{\endgroup\@href {#1}{\urlprefix }}%
\providecommand \urlprefix  [0]{URL }%
\providecommand \Eprint [0]{\href }%
\providecommand \doibase [0]{https://doi.org/}%
\providecommand \selectlanguage [0]{\@gobble}%
\providecommand \bibinfo  [0]{\@secondoftwo}%
\providecommand \bibfield  [0]{\@secondoftwo}%
\providecommand \translation [1]{[#1]}%
\providecommand \BibitemOpen [0]{}%
\providecommand \bibitemStop [0]{}%
\providecommand \bibitemNoStop [0]{.\EOS\space}%
\providecommand \EOS [0]{\spacefactor3000\relax}%
\providecommand \BibitemShut  [1]{\csname bibitem#1\endcsname}%
\let\auto@bib@innerbib\@empty
\bibitem [{\citenamefont {L\'opez}\ and\ \citenamefont {Valani}(2024)}]{lopez2024mega}%
  \BibitemOpen
  \bibfield  {author} {\bibinfo {author} {\bibfnamefont {A.~G.}\ \bibnamefont {L\'opez}}\ and\ \bibinfo {author} {\bibfnamefont {R.~N.}\ \bibnamefont {Valani}},\ }\bibfield  {title} {\bibinfo {title} {Megastable quantization in generalized pilot-wave hydrodynamics},\ }\href@noop {} {\bibfield  {journal} {\bibinfo  {journal} {arXiv:2410.12849}\ }\textbf {\bibinfo {volume} {[nlin.AO]}},\ \bibinfo {pages} {1} (\bibinfo {year} {2024})}\BibitemShut {NoStop}%
\bibitem [{\citenamefont {Kahn}\ and\ \citenamefont {Zarmi}(2014)}]{Kahn2014-na}%
  \BibitemOpen
  \bibfield  {author} {\bibinfo {author} {\bibfnamefont {P.~B.}\ \bibnamefont {Kahn}}\ and\ \bibinfo {author} {\bibfnamefont {Y.}~\bibnamefont {Zarmi}},\ }\href@noop {} {\emph {\bibinfo {title} {Nonlinear Dynamics}}},\ Dover Books on Physics\ (\bibinfo  {publisher} {Dover Publications},\ \bibinfo {address} {Mineola, NY},\ \bibinfo {year} {2014})\BibitemShut {NoStop}%
\bibitem [{\citenamefont {Zarmi}(2017)}]{ZARMI201721}%
  \BibitemOpen
  \bibfield  {author} {\bibinfo {author} {\bibfnamefont {Y.}~\bibnamefont {Zarmi}},\ }\bibfield  {title} {\bibinfo {title} {A classical limit-cycle system that mimics the quantum-mechanical harmonic oscillator},\ }\href {https://doi.org/https://doi.org/10.1016/j.physd.2017.08.003} {\bibfield  {journal} {\bibinfo  {journal} {Phys. D: Nonlinear Phenom.}\ }\textbf {\bibinfo {volume} {359}},\ \bibinfo {pages} {21} (\bibinfo {year} {2017})}\BibitemShut {NoStop}%
\bibitem [{\citenamefont {Raju}(2004)}]{raju2004electrodynamic}%
  \BibitemOpen
  \bibfield  {author} {\bibinfo {author} {\bibfnamefont {C.}~\bibnamefont {Raju}},\ }\bibfield  {title} {\bibinfo {title} {The electrodynamic 2-body problem and the origin of quantum mechanics},\ }\href@noop {} {\bibfield  {journal} {\bibinfo  {journal} {Foundations of Physics}\ }\textbf {\bibinfo {volume} {34}},\ \bibinfo {pages} {937} (\bibinfo {year} {2004})}\BibitemShut {NoStop}%
\bibitem [{\citenamefont {Schell}\ and\ \citenamefont {Ross}(1986)}]{schell1986effects}%
  \BibitemOpen
  \bibfield  {author} {\bibinfo {author} {\bibfnamefont {M.}~\bibnamefont {Schell}}\ and\ \bibinfo {author} {\bibfnamefont {J.}~\bibnamefont {Ross}},\ }\bibfield  {title} {\bibinfo {title} {Effects of time delay in rate processes},\ }\href@noop {} {\bibfield  {journal} {\bibinfo  {journal} {J. Chem. Phys}\ }\textbf {\bibinfo {volume} {85}},\ \bibinfo {pages} {6489} (\bibinfo {year} {1986})}\BibitemShut {NoStop}%
\bibitem [{\citenamefont {Airy}(1830)}]{airy1830certain}%
  \BibitemOpen
  \bibfield  {author} {\bibinfo {author} {\bibfnamefont {G.~B.}\ \bibnamefont {Airy}},\ }\href@noop {} {\emph {\bibinfo {title} {On certain Conditions under which a Perpetual Motion is possible}}}\ (\bibinfo  {publisher} {J. Smith},\ \bibinfo {year} {1830})\BibitemShut {NoStop}%
\bibitem [{\citenamefont {Mackey}\ and\ \citenamefont {Glass}(1977)}]{mackey1977oscillation}%
  \BibitemOpen
  \bibfield  {author} {\bibinfo {author} {\bibfnamefont {M.~C.}\ \bibnamefont {Mackey}}\ and\ \bibinfo {author} {\bibfnamefont {L.}~\bibnamefont {Glass}},\ }\bibfield  {title} {\bibinfo {title} {Oscillation and chaos in physiological control systems},\ }\href@noop {} {\bibfield  {journal} {\bibinfo  {journal} {Science}\ }\textbf {\bibinfo {volume} {197}},\ \bibinfo {pages} {287} (\bibinfo {year} {1977})}\BibitemShut {NoStop}%
\bibitem [{\citenamefont {Hansen}\ \emph {et~al.}(2022)\citenamefont {Hansen}, \citenamefont {Protachevicz}, \citenamefont {Iarosz}, \citenamefont {Caldas}, \citenamefont {Batista},\ and\ \citenamefont {Macau}}]{hansen2022effect}%
  \BibitemOpen
  \bibfield  {author} {\bibinfo {author} {\bibfnamefont {M.}~\bibnamefont {Hansen}}, \bibinfo {author} {\bibfnamefont {P.~R.}\ \bibnamefont {Protachevicz}}, \bibinfo {author} {\bibfnamefont {K.~C.}\ \bibnamefont {Iarosz}}, \bibinfo {author} {\bibfnamefont {I.~L.}\ \bibnamefont {Caldas}}, \bibinfo {author} {\bibfnamefont {A.~M.}\ \bibnamefont {Batista}},\ and\ \bibinfo {author} {\bibfnamefont {E.~E.}\ \bibnamefont {Macau}},\ }\bibfield  {title} {\bibinfo {title} {The effect of time delay for synchronisation suppression in neuronal networks},\ }\href@noop {} {\bibfield  {journal} {\bibinfo  {journal} {Chaos, Solit. Fractals}\ }\textbf {\bibinfo {volume} {164}},\ \bibinfo {pages} {112690} (\bibinfo {year} {2022})}\BibitemShut {NoStop}%
\bibitem [{\citenamefont {Ferrell}\ \emph {et~al.}(2011)\citenamefont {Ferrell}, \citenamefont {Tsai},\ and\ \citenamefont {Yang}}]{ferrell2011modeling}%
  \BibitemOpen
  \bibfield  {author} {\bibinfo {author} {\bibfnamefont {J.~E.}\ \bibnamefont {Ferrell}}, \bibinfo {author} {\bibfnamefont {T.~Y.-C.}\ \bibnamefont {Tsai}},\ and\ \bibinfo {author} {\bibfnamefont {Q.}~\bibnamefont {Yang}},\ }\bibfield  {title} {\bibinfo {title} {Modeling the cell cycle: why do certain circuits oscillate?},\ }\href@noop {} {\bibfield  {journal} {\bibinfo  {journal} {Cell}\ }\textbf {\bibinfo {volume} {144}},\ \bibinfo {pages} {874} (\bibinfo {year} {2011})}\BibitemShut {NoStop}%
\bibitem [{\citenamefont {Boutle}\ \emph {et~al.}(2007)\citenamefont {Boutle}, \citenamefont {Taylor},\ and\ \citenamefont {R{\"o}mer}}]{boutle2007nino}%
  \BibitemOpen
  \bibfield  {author} {\bibinfo {author} {\bibfnamefont {I.}~\bibnamefont {Boutle}}, \bibinfo {author} {\bibfnamefont {R.~H.}\ \bibnamefont {Taylor}},\ and\ \bibinfo {author} {\bibfnamefont {R.~A.}\ \bibnamefont {R{\"o}mer}},\ }\bibfield  {title} {\bibinfo {title} {El ni{\~n}o and the delayed action oscillator},\ }\href@noop {} {\bibfield  {journal} {\bibinfo  {journal} {Am. J. Phys.}\ }\textbf {\bibinfo {volume} {75}},\ \bibinfo {pages} {15} (\bibinfo {year} {2007})}\BibitemShut {NoStop}%
\bibitem [{\citenamefont {Salpeter}\ and\ \citenamefont {Salpeter}(1998)}]{salpeter1998mathematical}%
  \BibitemOpen
  \bibfield  {author} {\bibinfo {author} {\bibfnamefont {E.~E.}\ \bibnamefont {Salpeter}}\ and\ \bibinfo {author} {\bibfnamefont {S.~R.}\ \bibnamefont {Salpeter}},\ }\bibfield  {title} {\bibinfo {title} {Mathematical model for the epidemiology of tuberculosis, with estimates of the reproductive number and infection-delay function},\ }\href@noop {} {\bibfield  {journal} {\bibinfo  {journal} {Am. J. Epidemiol.}\ }\textbf {\bibinfo {volume} {147}},\ \bibinfo {pages} {398} (\bibinfo {year} {1998})}\BibitemShut {NoStop}%
\bibitem [{\citenamefont {L\'{o}pez}(2023)}]{LOPEZ2023113412}%
  \BibitemOpen
  \bibfield  {author} {\bibinfo {author} {\bibfnamefont {A.~G.}\ \bibnamefont {L\'{o}pez}},\ }\bibfield  {title} {\bibinfo {title} {Orbit quantization in a retarded harmonic oscillator},\ }\href@noop {} {\bibfield  {journal} {\bibinfo  {journal} {Chaos Solit. Fractals}\ }\textbf {\bibinfo {volume} {170}},\ \bibinfo {pages} {113412} (\bibinfo {year} {2023})}\BibitemShut {NoStop}%
\bibitem [{\citenamefont {Couder}\ \emph {et~al.}(2005)\citenamefont {Couder}, \citenamefont {Fort}, \citenamefont {Gautier},\ and\ \citenamefont {Boudaoud}}]{Couder2005}%
  \BibitemOpen
  \bibfield  {author} {\bibinfo {author} {\bibfnamefont {Y.}~\bibnamefont {Couder}}, \bibinfo {author} {\bibfnamefont {E.}~\bibnamefont {Fort}}, \bibinfo {author} {\bibfnamefont {C.-H.}\ \bibnamefont {Gautier}},\ and\ \bibinfo {author} {\bibfnamefont {A.}~\bibnamefont {Boudaoud}},\ }\bibfield  {title} {\bibinfo {title} {From bouncing to floating: noncoalescence of drops on a fluid bath},\ }\href@noop {} {\bibfield  {journal} {\bibinfo  {journal} {Phys. Rev. Lett.}\ }\textbf {\bibinfo {volume} {94}},\ \bibinfo {pages} {177801} (\bibinfo {year} {2005})}\BibitemShut {NoStop}%
\bibitem [{\citenamefont {Perrard}\ \emph {et~al.}(2014)\citenamefont {Perrard}, \citenamefont {Labousse}, \citenamefont {Miskin}, \citenamefont {Fort},\ and\ \citenamefont {Couder}}]{Perrard2014a}%
  \BibitemOpen
  \bibfield  {author} {\bibinfo {author} {\bibfnamefont {S.}~\bibnamefont {Perrard}}, \bibinfo {author} {\bibfnamefont {M.}~\bibnamefont {Labousse}}, \bibinfo {author} {\bibfnamefont {M.}~\bibnamefont {Miskin}}, \bibinfo {author} {\bibfnamefont {E.}~\bibnamefont {Fort}},\ and\ \bibinfo {author} {\bibfnamefont {Y.}~\bibnamefont {Couder}},\ }\bibfield  {title} {\bibinfo {title} {Self-organization into quantized eigenstates of a classical wave-driven particle},\ }\href@noop {} {\bibfield  {journal} {\bibinfo  {journal} {Nat. Commun.}\ }\textbf {\bibinfo {volume} {5}},\ \bibinfo {pages} {3219} (\bibinfo {year} {2014})}\BibitemShut {NoStop}%
\bibitem [{\citenamefont {Bush}\ and\ \citenamefont {Oza}(2020)}]{Bush2020review}%
  \BibitemOpen
  \bibfield  {author} {\bibinfo {author} {\bibfnamefont {J.~W.~M.}\ \bibnamefont {Bush}}\ and\ \bibinfo {author} {\bibfnamefont {A.~U.}\ \bibnamefont {Oza}},\ }\bibfield  {title} {\bibinfo {title} {Hydrodynamic quantum analogs},\ }\href@noop {} {\bibfield  {journal} {\bibinfo  {journal} {Rep. Prog. Phys.}\ }\textbf {\bibinfo {volume} {84}},\ \bibinfo {pages} {017001} (\bibinfo {year} {2020})}\BibitemShut {NoStop}%
\bibitem [{\citenamefont {Sprott}\ \emph {et~al.}(2017)\citenamefont {Sprott}, \citenamefont {Jafari}, \citenamefont {Khalaf},\ and\ \citenamefont {Kapitaniak}}]{Sprott2017}%
  \BibitemOpen
  \bibfield  {author} {\bibinfo {author} {\bibfnamefont {J.~C.}\ \bibnamefont {Sprott}}, \bibinfo {author} {\bibfnamefont {S.}~\bibnamefont {Jafari}}, \bibinfo {author} {\bibfnamefont {A.~J.~M.}\ \bibnamefont {Khalaf}},\ and\ \bibinfo {author} {\bibfnamefont {T.}~\bibnamefont {Kapitaniak}},\ }\bibfield  {title} {\bibinfo {title} {Megastability: Coexistence of a countable infinity of nested attractors in a periodically-forced oscillator with spatially-periodic damping},\ }\href@noop {} {\bibfield  {journal} {\bibinfo  {journal} {Eur. Phys. J. Spec. Top.}\ }\textbf {\bibinfo {volume} {226}},\ \bibinfo {pages} {1979} (\bibinfo {year} {2017})}\BibitemShut {NoStop}%
\bibitem [{\citenamefont {Álvaro G.~López}\ and\ \citenamefont {Valani}(2024)}]{Lopez2024selfexcited}%
  \BibitemOpen
  \bibfield  {author} {\bibinfo {author} {\bibnamefont {Álvaro G.~López}}\ and\ \bibinfo {author} {\bibfnamefont {R.~N.}\ \bibnamefont {Valani}},\ }\href@noop {} {\bibinfo {title} {Megastable quantization in self-excited systems}} (\bibinfo {year} {2024}),\ \Eprint {https://arxiv.org/abs/arXiv:2406.03906} {arXiv:2406.03906} \BibitemShut {NoStop}%
\bibitem [{\citenamefont {Grines}\ \emph {et~al.}(2018)\citenamefont {Grines}, \citenamefont {Osipov},\ and\ \citenamefont {Pikovsky}}]{grines2018describing}%
  \BibitemOpen
  \bibfield  {author} {\bibinfo {author} {\bibfnamefont {E.}~\bibnamefont {Grines}}, \bibinfo {author} {\bibfnamefont {G.}~\bibnamefont {Osipov}},\ and\ \bibinfo {author} {\bibfnamefont {A.}~\bibnamefont {Pikovsky}},\ }\bibfield  {title} {\bibinfo {title} {Describing dynamics of driven multistable oscillators with phase transfer curves},\ }\href@noop {} {\bibfield  {journal} {\bibinfo  {journal} {Chaos}\ }\textbf {\bibinfo {volume} {28}} (\bibinfo {year} {2018})}\BibitemShut {NoStop}%
\bibitem [{\citenamefont {Ahmadi}\ \emph {et~al.}(2007)\citenamefont {Ahmadi}, \citenamefont {Michalska},\ and\ \citenamefont {Buehler}}]{4252155}%
  \BibitemOpen
  \bibfield  {author} {\bibinfo {author} {\bibfnamefont {M.}~\bibnamefont {Ahmadi}}, \bibinfo {author} {\bibfnamefont {H.}~\bibnamefont {Michalska}},\ and\ \bibinfo {author} {\bibfnamefont {M.}~\bibnamefont {Buehler}},\ }\bibfield  {title} {\bibinfo {title} {Control and stability analysis of limit cycles in a hopping robot},\ }\href {https://doi.org/10.1109/TRO.2007.898956} {\bibfield  {journal} {\bibinfo  {journal} {IEEE Transactions on Robotics}\ }\textbf {\bibinfo {volume} {23}},\ \bibinfo {pages} {553} (\bibinfo {year} {2007})}\BibitemShut {NoStop}%
\bibitem [{\citenamefont {Li\'enard}(1898)}]{lien98}%
  \BibitemOpen
  \bibfield  {author} {\bibinfo {author} {\bibfnamefont {A.}~\bibnamefont {Li\'enard}},\ }\bibfield  {title} {\bibinfo {title} {Champ électrique et magnétique produit par une charge concentrée en un point et animée d'un mouvement quelconque},\ }\href@noop {} {\bibfield  {journal} {\bibinfo  {journal} {L'Éclairage Électrique}\ }\textbf {\bibinfo {volume} {16}},\ \bibinfo {pages} {5} (\bibinfo {year} {1898})}\BibitemShut {NoStop}%
\bibitem [{\citenamefont {L\'{o}pez}(2020)}]{onanelec}%
  \BibitemOpen
  \bibfield  {author} {\bibinfo {author} {\bibfnamefont {A.~G.}\ \bibnamefont {L\'{o}pez}},\ }\bibfield  {title} {\bibinfo {title} {On an electrodynamic origin of quantum fluctuations},\ }\href@noop {} {\bibfield  {journal} {\bibinfo  {journal} {Nonlinear Dyn.}\ }\textbf {\bibinfo {volume} {102}},\ \bibinfo {pages} {621} (\bibinfo {year} {2020})}\BibitemShut {NoStop}%
\bibitem [{\citenamefont {López}\ and\ \citenamefont {Valani}(2024)}]{lopval24}%
  \BibitemOpen
  \bibfield  {author} {\bibinfo {author} {\bibfnamefont {A.~G.}\ \bibnamefont {López}}\ and\ \bibinfo {author} {\bibfnamefont {R.~N.}\ \bibnamefont {Valani}},\ }\bibfield  {title} {\bibinfo {title} {Unpredictable tunneling in a retarded bistable potential},\ }\href@noop {} {\bibfield  {journal} {\bibinfo  {journal} {Chaos}\ }\textbf {\bibinfo {volume} {34}},\ \bibinfo {pages} {043117} (\bibinfo {year} {2024})}\BibitemShut {NoStop}%
\bibitem [{\citenamefont {Jenkins}(2013)}]{jenkins}%
  \BibitemOpen
  \bibfield  {author} {\bibinfo {author} {\bibfnamefont {A.}~\bibnamefont {Jenkins}},\ }\bibfield  {title} {\bibinfo {title} {Self-oscillation},\ }\href@noop {} {\bibfield  {journal} {\bibinfo  {journal} {Phys. Rep.}\ }\textbf {\bibinfo {volume} {525}},\ \bibinfo {pages} {167} (\bibinfo {year} {2013})}\BibitemShut {NoStop}%
\bibitem [{\citenamefont {Shampine}(2005)}]{shampine05}%
  \BibitemOpen
  \bibfield  {author} {\bibinfo {author} {\bibfnamefont {L.}~\bibnamefont {Shampine}},\ }\bibfield  {title} {\bibinfo {title} {Solving {ODEs} and {DDEs} with residual control},\ }\href@noop {} {\bibfield  {journal} {\bibinfo  {journal} {Appl. Numer. Math.}\ }\textbf {\bibinfo {volume} {52}},\ \bibinfo {pages} {113} (\bibinfo {year} {2005})}\BibitemShut {NoStop}%
\bibitem [{\citenamefont {Erneux}\ \emph {et~al.}(2023)\citenamefont {Erneux}, \citenamefont {Kovalev},\ and\ \citenamefont {Viktorov}}]{Erneu23}%
  \BibitemOpen
  \bibfield  {author} {\bibinfo {author} {\bibfnamefont {T.}~\bibnamefont {Erneux}}, \bibinfo {author} {\bibfnamefont {A.~V.}\ \bibnamefont {Kovalev}},\ and\ \bibinfo {author} {\bibfnamefont {E.~A.}\ \bibnamefont {Viktorov}},\ }\bibfield  {title} {\bibinfo {title} {Short delay limit of the delayed duffing oscillator},\ }\href@noop {} {\bibfield  {journal} {\bibinfo  {journal} {Phys. Rev. E}\ }\textbf {\bibinfo {volume} {108}},\ \bibinfo {pages} {064201} (\bibinfo {year} {2023})}\BibitemShut {NoStop}%
\bibitem [{\citenamefont {Krylov}\ and\ \citenamefont {Bogoliubov}(1950)}]{krylov1950}%
  \BibitemOpen
  \bibfield  {author} {\bibinfo {author} {\bibfnamefont {N.~M.}\ \bibnamefont {Krylov}}\ and\ \bibinfo {author} {\bibfnamefont {N.~N.}\ \bibnamefont {Bogoliubov}},\ }\href@noop {} {\emph {\bibinfo {title} {Introduction to non-linear mechanics}}},\ \bibinfo {number} {11}\ (\bibinfo  {publisher} {Princeton university press},\ \bibinfo {year} {1950})\BibitemShut {NoStop}%
\bibitem [{\citenamefont {Bohm}(1952)}]{bohm52}%
  \BibitemOpen
  \bibfield  {author} {\bibinfo {author} {\bibfnamefont {D.}~\bibnamefont {Bohm}},\ }\bibfield  {title} {\bibinfo {title} {A suggested interpretation of the quantum theory in terms of ``hidden" variables. {I}},\ }\href@noop {} {\bibfield  {journal} {\bibinfo  {journal} {Phys. Rev.}\ }\textbf {\bibinfo {volume} {85}},\ \bibinfo {pages} {166} (\bibinfo {year} {1952})}\BibitemShut {NoStop}%
\bibitem [{Note1()}]{Note1}%
  \BibitemOpen
  \bibinfo {note} {Note, we could have used $k$ instead of $k+\alpha $, regarding to the original Eq.~\protect \eqref {eq:1}, as well. However, we have considered Eq.~\protect \eqref {eq:3}, instead. A careful consideration hints at the use of the latter, since the delay term in Eq.~\protect \eqref {eq:1} includes a conservative component when this term is expanded in a Taylor series, and it relates to the zeroth-order contribution (see Ref.~\cite {Davidow2017}). Thus, we use $k+\alpha $, which gives more accurate fitting results.}\BibitemShut {Stop}%
\bibitem [{\citenamefont {López}\ \emph {et~al.}(2023)\citenamefont {López}, \citenamefont {Benito}, \citenamefont {Sabuco},\ and\ \citenamefont {Delgado-Bonal}}]{lopezte}%
  \BibitemOpen
  \bibfield  {author} {\bibinfo {author} {\bibfnamefont {A.~G.}\ \bibnamefont {López}}, \bibinfo {author} {\bibfnamefont {F.}~\bibnamefont {Benito}}, \bibinfo {author} {\bibfnamefont {J.}~\bibnamefont {Sabuco}},\ and\ \bibinfo {author} {\bibfnamefont {A.}~\bibnamefont {Delgado-Bonal}},\ }\bibfield  {title} {\bibinfo {title} {The thermodynamic efficiency of the {Lorenz} system},\ }\href@noop {} {\bibfield  {journal} {\bibinfo  {journal} {Chaos, Solit. Fractals}\ }\textbf {\bibinfo {volume} {172}},\ \bibinfo {pages} {113521} (\bibinfo {year} {2023})}\BibitemShut {NoStop}%
\bibitem [{\citenamefont {Mackey}(2011)}]{mackey2011}%
  \BibitemOpen
  \bibfield  {author} {\bibinfo {author} {\bibfnamefont {M.~C.}\ \bibnamefont {Mackey}},\ }\href@noop {} {\emph {\bibinfo {title} {Time's Arrow: The origins of thermodynamic behavior}}}\ (\bibinfo  {publisher} {Courier Corporation},\ \bibinfo {year} {2011})\BibitemShut {NoStop}%
\bibitem [{\citenamefont {Prigogine}(1978)}]{Pri78}%
  \BibitemOpen
  \bibfield  {author} {\bibinfo {author} {\bibfnamefont {I.}~\bibnamefont {Prigogine}},\ }\bibfield  {title} {\bibinfo {title} {Time, structure, and fluctuations},\ }\href@noop {} {\bibfield  {journal} {\bibinfo  {journal} {Science}\ }\textbf {\bibinfo {volume} {201}},\ \bibinfo {pages} {777} (\bibinfo {year} {1978})}\BibitemShut {NoStop}%
\bibitem [{\citenamefont {Davidow}\ \emph {et~al.}(2017)\citenamefont {Davidow}, \citenamefont {Shayak},\ and\ \citenamefont {Rand}}]{Davidow2017}%
  \BibitemOpen
  \bibfield  {author} {\bibinfo {author} {\bibfnamefont {M.}~\bibnamefont {Davidow}}, \bibinfo {author} {\bibfnamefont {B.}~\bibnamefont {Shayak}},\ and\ \bibinfo {author} {\bibfnamefont {R.~H.}\ \bibnamefont {Rand}},\ }\bibfield  {title} {\bibinfo {title} {Analysis of a remarkable singularity in a nonlinear {DDE}},\ }\href@noop {} {\bibfield  {journal} {\bibinfo  {journal} {Nonlinear Dyn.}\ }\textbf {\bibinfo {volume} {90}},\ \bibinfo {pages} {317} (\bibinfo {year} {2017})}\BibitemShut {NoStop}%
\bibitem [{\citenamefont {Pikovsky}\ \emph {et~al.}(2001)\citenamefont {Pikovsky}, \citenamefont {Rosenblum},\ and\ \citenamefont {Kurths}}]{Pikovsky_Rosenblum_Kurths_2001}%
  \BibitemOpen
  \bibfield  {author} {\bibinfo {author} {\bibfnamefont {A.}~\bibnamefont {Pikovsky}}, \bibinfo {author} {\bibfnamefont {M.}~\bibnamefont {Rosenblum}},\ and\ \bibinfo {author} {\bibfnamefont {J.}~\bibnamefont {Kurths}},\ }\href@noop {} {\emph {\bibinfo {title} {Synchronization: A Universal Concept in Nonlinear Sciences}}},\ Cambridge Nonlinear Science Series\ (\bibinfo  {publisher} {Cambridge University Press},\ \bibinfo {year} {2001})\BibitemShut {NoStop}%
\bibitem [{\citenamefont {Gandhimathi}\ \emph {et~al.}(2006)\citenamefont {Gandhimathi}, \citenamefont {Rajasekar},\ and\ \citenamefont {Kurths}}]{gandhimathi2006vibrational}%
  \BibitemOpen
  \bibfield  {author} {\bibinfo {author} {\bibfnamefont {V.}~\bibnamefont {Gandhimathi}}, \bibinfo {author} {\bibfnamefont {S.}~\bibnamefont {Rajasekar}},\ and\ \bibinfo {author} {\bibfnamefont {J.}~\bibnamefont {Kurths}},\ }\bibfield  {title} {\bibinfo {title} {Vibrational and stochastic resonances in two coupled overdamped anharmonic oscillators},\ }\href@noop {} {\bibfield  {journal} {\bibinfo  {journal} {Phys. Lett. A}\ }\textbf {\bibinfo {volume} {360}},\ \bibinfo {pages} {279} (\bibinfo {year} {2006})}\BibitemShut {NoStop}%
\bibitem [{Note2()}]{Note2}%
  \BibitemOpen
  \bibinfo {note} {Variations in $\phi $ seems to introduce an uncertainty of one energy level during transitions i.e. the final orbit after harmonic pulse is above or below one level. In fact, the effect of modifying $t_0$ is very similar to changing the value of $\phi $, and both parameters have a weak effect on transitions between orbits.}\BibitemShut {Stop}%
\bibitem [{\citenamefont {Han}\ \emph {et~al.}(2009)\citenamefont {Han}, \citenamefont {Yang},\ and\ \citenamefont {Yu}}]{doi:10.1142/S0218127409025250}%
  \BibitemOpen
  \bibfield  {author} {\bibinfo {author} {\bibfnamefont {M.}~\bibnamefont {Han}}, \bibinfo {author} {\bibfnamefont {J.}~\bibnamefont {Yang}},\ and\ \bibinfo {author} {\bibfnamefont {P.}~\bibnamefont {Yu}},\ }\bibfield  {title} {\bibinfo {title} {Hopf bifurcations for near-{H}amiltonian systems},\ }\href {https://doi.org/10.1142/S0218127409025250} {\bibfield  {journal} {\bibinfo  {journal} {Int. J. Bifurc. Chaos}\ }\textbf {\bibinfo {volume} {19}},\ \bibinfo {pages} {4117} (\bibinfo {year} {2009})}\BibitemShut {NoStop}%
\bibitem [{\citenamefont {Bohr}(1913)}]{bohr13}%
  \BibitemOpen
  \bibfield  {author} {\bibinfo {author} {\bibfnamefont {N.}~\bibnamefont {Bohr}},\ }\bibfield  {title} {\bibinfo {title} {I. {On} the constitution of atoms and molecules},\ }\href@noop {} {\bibfield  {journal} {\bibinfo  {journal} {The London, Edinburgh, and Dublin Philosophical Magazine and Journal of Science}\ }\textbf {\bibinfo {volume} {26}},\ \bibinfo {pages} {1} (\bibinfo {year} {1913})}\BibitemShut {NoStop}%
\bibitem [{\citenamefont {de~la Pe\~{n}a}\ \emph {et~al.}(2014)\citenamefont {de~la Pe\~{n}a}, \citenamefont {Cetto},\ and\ \citenamefont {Vald\'{e}s-Hernandes}}]{Pen14}%
  \BibitemOpen
  \bibfield  {author} {\bibinfo {author} {\bibfnamefont {L.}~\bibnamefont {de~la Pe\~{n}a}}, \bibinfo {author} {\bibfnamefont {A.~M.}\ \bibnamefont {Cetto}},\ and\ \bibinfo {author} {\bibfnamefont {A.}~\bibnamefont {Vald\'{e}s-Hernandes}},\ }\bibfield  {title} {\bibinfo {title} {The zero-point field and the emergence of the quantum},\ }\href@noop {} {\bibfield  {journal} {\bibinfo  {journal} {Int. J. Mod. Phys. E}\ }\textbf {\bibinfo {volume} {23}},\ \bibinfo {pages} {1450049} (\bibinfo {year} {2014})}\BibitemShut {NoStop}%
\bibitem [{\citenamefont {Nusse}\ and\ \citenamefont {Yorke}(1996)}]{nuss96}%
  \BibitemOpen
  \bibfield  {author} {\bibinfo {author} {\bibfnamefont {H.~E.}\ \bibnamefont {Nusse}}\ and\ \bibinfo {author} {\bibfnamefont {J.~A.}\ \bibnamefont {Yorke}},\ }\bibfield  {title} {\bibinfo {title} {Basins of attraction},\ }\href@noop {} {\bibfield  {journal} {\bibinfo  {journal} {Science}\ }\textbf {\bibinfo {volume} {271}},\ \bibinfo {pages} {1376} (\bibinfo {year} {1996})}\BibitemShut {NoStop}%
\bibitem [{\citenamefont {Van~der Pol}(1960)}]{van1960theory}%
  \BibitemOpen
  \bibfield  {author} {\bibinfo {author} {\bibfnamefont {B.}~\bibnamefont {Van~der Pol}},\ }\bibfield  {title} {\bibinfo {title} {A theory of the amplitude of free and forced triode vibrations, {R}adio {R}ev. 1 (1920) 701-710, 754-762; selected scientific papers, {V}ol. i},\ }\href@noop {} {\bibfield  {journal} {\bibinfo  {journal} {ed: North Holland}\ } (\bibinfo {year} {1960})}\BibitemShut {NoStop}%
\bibitem [{\citenamefont {Dunnmon}\ \emph {et~al.}(2011)\citenamefont {Dunnmon}, \citenamefont {Stanton}, \citenamefont {Mann},\ and\ \citenamefont {Dowell}}]{DUNNMON20111182}%
  \BibitemOpen
  \bibfield  {author} {\bibinfo {author} {\bibfnamefont {J.}~\bibnamefont {Dunnmon}}, \bibinfo {author} {\bibfnamefont {S.}~\bibnamefont {Stanton}}, \bibinfo {author} {\bibfnamefont {B.}~\bibnamefont {Mann}},\ and\ \bibinfo {author} {\bibfnamefont {E.}~\bibnamefont {Dowell}},\ }\bibfield  {title} {\bibinfo {title} {Power extraction from aeroelastic limit cycle oscillations},\ }\href {https://doi.org/https://doi.org/10.1016/j.jfluidstructs.2011.02.003} {\bibfield  {journal} {\bibinfo  {journal} {J. Fluids Struct.}\ }\textbf {\bibinfo {volume} {27}},\ \bibinfo {pages} {1182} (\bibinfo {year} {2011})}\BibitemShut {NoStop}%
\bibitem [{\citenamefont {Zhang}\ \emph {et~al.}(2024)\citenamefont {Zhang}, \citenamefont {Yan}, \citenamefont {Dong}, \citenamefont {Dykman},\ and\ \citenamefont {Chan}}]{PhysRevApplied.22.034072}%
  \BibitemOpen
  \bibfield  {author} {\bibinfo {author} {\bibfnamefont {B.}~\bibnamefont {Zhang}}, \bibinfo {author} {\bibfnamefont {Y.}~\bibnamefont {Yan}}, \bibinfo {author} {\bibfnamefont {X.}~\bibnamefont {Dong}}, \bibinfo {author} {\bibfnamefont {M.}~\bibnamefont {Dykman}},\ and\ \bibinfo {author} {\bibfnamefont {H.}~\bibnamefont {Chan}},\ }\bibfield  {title} {\bibinfo {title} {Frequency stabilization of self-sustained oscillations in a sideband-driven electromechanical resonator},\ }\href {https://doi.org/10.1103/PhysRevApplied.22.034072} {\bibfield  {journal} {\bibinfo  {journal} {Phys. Rev. Appl.}\ }\textbf {\bibinfo {volume} {22}},\ \bibinfo {pages} {034072} (\bibinfo {year} {2024})}\BibitemShut {NoStop}%
\bibitem [{\citenamefont {Xu}\ \emph {et~al.}(2022)\citenamefont {Xu}, \citenamefont {Krisnanda},\ and\ \citenamefont {Liew}}]{xu2022limit}%
  \BibitemOpen
  \bibfield  {author} {\bibinfo {author} {\bibfnamefont {X.}~\bibnamefont {Xu}}, \bibinfo {author} {\bibfnamefont {T.}~\bibnamefont {Krisnanda}},\ and\ \bibinfo {author} {\bibfnamefont {T.~C.}\ \bibnamefont {Liew}},\ }\bibfield  {title} {\bibinfo {title} {Limit cycles and chaos in the hybrid atom-optomechanics system},\ }\href@noop {} {\bibfield  {journal} {\bibinfo  {journal} {Sci. Rep.}\ }\textbf {\bibinfo {volume} {12}},\ \bibinfo {pages} {15288} (\bibinfo {year} {2022})}\BibitemShut {NoStop}%
\end{thebibliography}%

\end{document}